\documentclass[acmsmall,screen]{acmart}
\usepackage{graphicx} % Required for inserting images
\usepackage{listings} % Code listing rendering
\usepackage{algorithm}
\usepackage{algpseudocode}
\usepackage{multirow}
\usepackage{multicol}
\usepackage{subcaption}
\usepackage{xspace}
\usepackage[inline]{enumitem}
\usepackage{tikz}
\usepackage{hyperref}
\usetikzlibrary{arrows.meta,calc}
\usepackage{booktabs,tabularx,array}

\newcolumntype{Y}[1]{>{\raggedright\arraybackslash\hsize=#1\hsize}X}
\usepackage{xcolor}

\makeatletter
\let\ALG@step@orig\ALG@step
\def\ALG@step{%
  \refstepcounter{ALG@line}%   % increment line number and create ALG@line.<alg>.<line> anchor
  \addtocounter{ALG@rem}{1}%
  \ifthenelse{\equal{\arabic{ALG@rem}}{\ALG@numberfreq}}%
     {\setcounter{ALG@rem}{0}\alglinenumber{\arabic{ALG@line}}}%
     {}%
}
\renewcommand{\theHALG@line}{\arabic{algorithm}.\arabic{ALG@line}}
\def\algref#1#2{\hyperref[#2]{Algorithm~\ref*{#1} (line~\ref*{#2})}}
\makeatother
\title{Pattern-Guided Graph Synthesis for Suppressing Known Defects in DL Compiler Fuzzing}

\newcommand{\numBugs}{25}
\newcommand{\numCrashes}{15}
\newcommand{\numWrongCode}{10}

\newcommand{\numFixed}{7}
\newcommand{\numConfirmed}{12}
\newcommand{\numReported}{5}
\newcommand{\numNotPlanned}{1}
\newcommand{\numFixedConfirmed}{19}
\newcommand{\toolName}{\textsc{Reprise}}
\newcommand{\tvmIssue}[1]{\href{https://github.com/apache/tvm/issues/#1}{TVM-#1}}
\newcommand{\ptIssue}[1]{\href{https://github.com/pytorch/pytorch/issues/#1}{PYTORCH-#1}}
\newcommand{\ortIssue}[1]{\href{https://github.com/microsoft/onnxruntime/issues/#1}{ORT-#1}}

\newcommand{\nRepeats}{5\xspace}

\begin{document}

\author{Qiong Feng}
\orcid{0000-0003-1667-8062}
\affiliation{%
  \institution{School of Computer Science and Engineering, Nanjing University of Science and Technology}
  %\streetaddress{NO. 152 Luoyu Road}
  %\city{Wuhan}
  %\state{Hubei}
  \country{China}
}
\email{qiongfeng@njust.edu.cn}

\author{Xiaotian Ma}
\orcid{0009-0008-9121-3656}
\affiliation{%
  \institution{School of Computer Science and Engineering, Nanjing University of Science and Technology}
  %\streetaddress{NO. 152 Luoyu Road}
  %\city{Wuhan}
  %\state{Hubei}
  \country{China}
}
\email{xyzboom@njust.edu.cn}

\author{Di Cui}
\orcid{0000-0001-5056-2704}
\affiliation{%
  \institution{Department of Computer Science, Xidian University}
  %\streetaddress{NO. 152 Luoyu Road}
  %\city{Wuhan}
  %\state{Hubei}
  \country{China}
}
\email{cuidi@xidian.edu.cn}

\author{Wei Song}
\orcid{0000-0002-4324-3382}
\affiliation{%
  \institution{School of Computer Science and Engineering, Nanjing University of Science and Technology}
  %\streetaddress{NO. 1037 Luoyu Road}
  %\city{Wuhan}
  %\state{Hubei}
  \country{China}
}
\email{wsong@njust.edu.cn}

\author{Peng Liang}
\orcid{0000-0002-2056-5346}
%\authornote{Corresponding author.}
\affiliation{%
  \institution{School of Computer Science, Wuhan University}
  %\streetaddress{NO. 299 Bayi Road}
  %\city{Wuhan}
  %\state{Hubei}
  \country{China}
}
\email{liangp@whu.edu.cn}

\begin{abstract}
Fuzzing is effective at finding bugs in deep learning (DL) compilers, but existing fuzzers repeatedly trigger faults they have already uncovered.
Current fuzzers diversify the generated programs, and downstream tools deduplicate bug reports post hoc, but neither stops a fuzzer from generating programs that re-trigger known defects.
We present \toolName{}, a DL compiler fuzzer that suppresses reports of known defects during test generation.
\toolName{} uses a deliberately lightweight generator that builds graphs in a unified intermediate representation (UIR) with local operator signatures.
Instead of diversifying programs, \toolName{} distills each discovered defect into a semantic graph pattern that captures the operators, value constraints, graph context, and data flow required to trigger it.
During graph synthesis, it regenerates any node that completes a known pattern, so known defect triggers are avoided before compilation and execution.
We evaluate \toolName{} on three DL compilers: TVM, PyTorch Inductor, and ONNX Runtime.
Against its unguided variant, patterns written for the dominant known defects reduce crash reports by 90.2\% on TVM (254 to 25) and by 93.8\% on PyTorch Inductor (594 to 37), with comparable branch coverage and 1.0--17.9\% fewer executed tests.
These results come from one run per configuration and concern the defects the patterns were distilled from.
\toolName{} also found \numBugs{} previously unknown bugs across the three compilers, 24 of them within one month, of which \numFixed{} have been fixed and \numConfirmed{} more confirmed by developers.
The replication package has been provided at~\cite{dataset}.
\end{abstract}
\maketitle

\begin{comment}
    Discussion
related work add Add white fox

Add discussion how pattern can be used for generation code fuzzing
Pattern融合

Threats
一个pattern对于不同的bug

APPRoach Pattern宽窄

RQ1. 带有去重策略的AIFuzzer的效果：覆盖率和 其他SOTA是不是有损失？能不能达到SOTA结果？
1.1
1.2    消融：pattern过宽 没bug
RQ2. 消融：去重模块。。。去重降低了多少人工成本？带有5个pattern降低了
	  
RQ2. How does AIFuzzer compare against state of the art in DL compiler fuzzing in terms of code coverage?
RQ3. an AIFuzzer find new bugs?
\end{comment}

\section{Introduction}
\label{sec:introduction}

Deep Learning (DL) compilers have become critical software infrastructure for
optimizing neural network models across heterogeneous hardware backends.
Their complex graph transformations and code generation pipelines make
automated testing and fuzzing increasingly important. Generation-based
fuzzers synthesize computation graphs from operator specifications
~\cite{liu2023nnsmith,liu2023neuri,wang2023gencog,ma2023fuzzing}, while
mutation-based fuzzers derive new programs from existing ones
~\cite{liu2022coverage,wei2022free}. These techniques have exposed many
compiler bugs, but they also repeatedly trigger faults that have already
been discovered. Random generation tends to exercise some faulty code paths
much more often than others~\cite{chen2013taming}, so the same underlying
defect is triggered by many different programs. Such redundant findings add
little information to new defects, yet developers must still reproduce, minimize, classify,
and triage each failure. Such repeatedly triggered faults also consume testing effort that could be used to explore new behaviors.

Existing fuzzing techniques primarily optimize the diversity or coverage of
the \emph{generated programs}, rather than the diversity of the \emph{bugs
they trigger}. NNSmith diversifies graph topologies and operator attributes
~\cite{liu2023nnsmith}, GenCoG maximizes vertex and edge diversity
~\cite{wang2023gencog}, and HirGen uses coverage criteria to guide graph
generation~\cite{ma2023fuzzing}. Tzer~\cite{liu2022coverage} and
FUEL~\cite{yang2025may} retain inputs that exercise new code. These
strategies encourage exploration of different programs or code regions, but
they do not recognize that different programs may trigger the same defect.
Program diversity therefore does not necessarily translate into diversity of
discovered defects. Downstream deduplication tools group duplicate reports
only after the fuzzer has generated and executed the programs.

This gap suggests a different question: \emph{Can previously observed
bug-triggering structures be used to guide subsequent generation away from
redundant failures?} Although two programs may differ syntactically, their
failures can share characteristic structural configurations, such as
operator combinations, graph topology, or interactions among operators.
We therefore summarize previously confirmed bug-triggering programs as
semantic patterns and use these patterns as \emph{negative guidance} during
graph synthesis. When a newly generated node would complete a known
bug-triggering pattern, the fuzzer regenerates that node before the program
is compiled and executed. Our motivation is that effort no longer spent
re-triggering known defects could instead go toward previously untested
regions of the graph space.

To this end, we present \toolName{}, a DL compiler fuzzer built upon this insight. Its
generator is deliberately lightweight: it builds graphs in a unified
intermediate representation (UIR) with local operator signatures, without SMT
solving and without a program-diversification strategy. Instead of
diversifying programs, \toolName{} uses patterns of already-found defects to
reduce repeated bug-triggering executions. Patterns are currently written by
hand from reduced reproducers. The goal is not to suppress failures in
general, but to suppress reports of known defects.

We evaluate \toolName{} on three DL compilers: TVM, PyTorch Inductor, and ONNX Runtime. Against
its unguided variant, patterns written for the dominant known defects reduce
crash reports on TVM (CUDA) by 88.2\% with the patterns for the top defect
alone and by 90.2\% with the full set (254 to 25 reports over 10{,}000
programs), and on PyTorch Inductor by 93.8\% (594 to 37). Guidance leaves
branch coverage comparable to the unguided variant and executes 1.0--17.9\%
fewer tests within a fixed budget. These results come from one run per
configuration and concern the defects from which the patterns were distilled.
Compared with NeuRI~\cite{liu2023neuri}, a state-of-the-art fuzzer, \toolName{} covers more
branches on TVM but fewer on PyTorch Inductor and ONNX Runtime. In a
one-month fuzzing, \toolName{} found 24 of all \numBugs{} previously unknown
bugs across the three compilers, \numFixed{} of which have been fixed and
\numConfirmed{} more confirmed by developers. In summary, our key \textbf{contributions} are:

\begin{itemize}[leftmargin=*]
    \item \textbf{Redundancy study:} We quantify redundancy in NeuRI's crash
    reports on TVM, PyTorch Inductor, and ONNX Runtime using report-weighted
    concentration measures from ecology and statistics. A single defect
    accounts for 64.6\%--98.6\% of the reports.
    \item \textbf{Pattern-guided graph synthesis:} We propose \toolName{}, a
    fuzzer that distills discovered defects into semantic graph patterns and
    uses these patterns to regenerate matching nodes during graph synthesis, on top of a
    lightweight, solver-free generator.
    \item \textbf{Evaluation and practical impact:} Patterns for dominant known
    defects cut crash reports by 90.2\% on TVM and 93.8\% on PyTorch Inductor
    relative to the unguided variant, with comparable coverage and a
    throughput cost. \toolName{} found \numBugs{} previously
    unknown bugs.
\end{itemize}

\section{Motivation}
\label{sec:motivation}

\subsection{Fuzzers Repeatedly Rediscover the Same Defects}
\label{sec:motivation-redundancy}

\begin{table}[t]
\centering
\caption{Crash reports and distinct bugs in each 10{,}000-program run generated by NeuRI.}
\label{tab:motivation-bugs}
\footnotesize
\setlength{\tabcolsep}{3pt}
\begin{tabularx}{\textwidth}{lllXrr}
\toprule
Platform & Distinct Bugs & Category & Symptom & \# Reports & Ratio \\
\midrule
\multirow{4}{*}{\shortstack[l]{TVM\\(CUDA)}}
& TVM-D1 & Schedule Bind & \texttt{ScheduleError}: dlight thread bind fail & 82 & 64.6\% \\
& TVM-D2 & Loop Fusion & \texttt{ScheduleError}: loops cannot be fused & 23 & 18.1\% \\
& TVM-D3 & Buffer Index & \texttt{ScheduleError}: reduction vars indexing & 19 & 15.0\% \\
& TVM-D4 & Rule Assert & \texttt{InternalError}: dlight GEMV assert & 3 & 2.4\% \\
\cmidrule{2-6}
& \emph{TVM Total ($U=4$)} & & & \emph{127} & \emph{100.0\%} \\
\midrule
\multirow{2}{*}{\shortstack[l]{ONNX Runtime\\(CPU)}}
& ORT-D1 & Missing Op & \texttt{OrtFail: NotImplemented}: double \texttt{Erf} & 272 & 98.6\% \\
& ORT-D2 & Graph Fusion & \texttt{OrtFail: InvalidGraph}: clip min dtype & 4 & 1.4\% \\
\cmidrule{2-6}
& \emph{ONNX Total ($U=2$)} & & & \emph{276} & \emph{100.0\%} \\
\midrule
\multirow{3}{*}{\shortstack[l]{PyTorch Inductor\\(CPU)}}
& PT-D1 & CodeGen Temp. & \texttt{CppCompileError}: \texttt{min/max\_propagate\_nan} & 40 & 81.6\% \\
& PT-D2 & Symbolic Trace & \texttt{InternalTorchDynamoError}: generator obj. & 5 & 10.2\% \\
& PT-D3 & Runtime Abort & Memory safety abort (\texttt{SIGABRT}) \& stride & 4 & 8.2\% \\
\cmidrule{2-6}
& \emph{PyTorch Total ($U=3$)} & & & \emph{49} & \emph{100.0\%} \\
\bottomrule
\end{tabularx}
%\vspace{2pt}
%{\footnotesize $^\dagger$Root-cause defect clusters identified during the 10{,}000-sample run; \tvmIssue{20048} and \tvmIssue{20047} were submitted to and tracked by upstream maintainers (Table~\ref{tab:bugs-detail}).}
\end{table}

To quantify how frequently DL compiler fuzzers rediscover known defects, we run
NeuRI~\cite{liu2023neuri}, a state-of-the-art program generator, to produce
10{,}000 programs for each of TVM, PyTorch Inductor, and ONNX Runtime. We
analyze only crash reports as their error messages and stack traces expose the
failing compiler component, which makes grouping by root cause reliable. Particularly, one author clusters the crash reports by root-cause
defect, a second author verifies the assignments, and disagreements are
resolved through discussion. We count distinct defects by root cause rather
than by heuristic crash signatures, which can substantially overcount or
undercount bugs~\cite{klees2018evaluating}. Table~\ref{tab:motivation-bugs} lists the
resulting defects and the number of reports each one caused.
 
Following the view of fuzzing as species discovery~\cite{bohme2018stads}, we
apply standard measures of concentration from ecology and statistics to bug reports 
and treat each defect as a species and each bug report as an individual sampled from it. 
Let $\mathcal{D}$ be the set of distinct defects found in a run,
$b_d$ the number of bug reports caused by defect $d \in \mathcal{D}$,
$B = \sum_d b_d$, and $U = |\mathcal{D}|$.
A natural measure of redundancy is the average number of reports per defect,
$B/U$. However, this average treats every defect equally, while developers
triage reports one at a time. On TVM, for example, NeuRI's 127 reports
correspond to 4 defects, so $B/U = 31.8$ (Table~\ref{tab:motivation-bugs}).
Yet, 82 of the reports belong to a single defect, so a randomly picked report
belongs to that defect with a probability of 64.6\%(82/127), and that defect has 82
reports, not 32. $B/U$ therefore understates the redundancy developers
actually face. This is a known effect of averaging group sizes over groups
rather than over their members~\cite{feld1977classsize}, and the gap grows as
reports concentrate on fewer defects. We instead average over reports: for
each report, we count the reports that share its root cause, including
itself, and take the mean:
\[
  R_w = \frac{1}{B} \sum_{d \in \mathcal{D}} b_d \cdot b_d
      = \frac{\sum_{d \in \mathcal{D}} b_d^2}{B}.
\]
Each defect $d$ has $b_d$ reports. On TVM, $R_w = 60.0$: a randomly picked report shares its root
cause with 59 others on average. $R_w$ equals $B / {}^2\!D$, where
${}^2\!D = B^2 / \sum_d b_d^2$ is the Hill number of order~2, also called the
\emph{effective number} of defects~\cite{hill1973diversity,jost2006entropy}.
This is the number of equally frequent defects that would produce the same
concentration of reports. $R_w$ thus replaces the raw defect count $U$ in
$B/U$ with the effective count ${}^2\!D \leq U$, so $R_w \geq B/U$, with
equality only when all defects are reported equally often. For NeuRI on TVM,
${}^2\!D = 2.1$: its four defects concentrate reports as if there were only
about two.
Since $R_w$ grows with the number of reports, we also use the scale-free
\emph{duplicate-pair rate}
\[
  C = \frac{R_w - 1}{B - 1} = \frac{\sum_{d} b_d (b_d - 1)}{B (B - 1)},
\]
which is Simpson's index of concentration~\cite{simpson1949measurement}: the
probability that two distinct, randomly chosen reports share a root cause.
Because it squares each defect's abundance, Simpson's index is dominated by
the most abundant categories, which suits redundancy caused by frequently
triggered defects.

\begin{table}[t]
\centering
\caption{Redundancy of crash reports in 10{,}000-program NeuRI runs.
$B$: crash reports; $U$: distinct defects; ${}^2\!D$: effective number of
defects; $R_w$: report-weighted redundancy; $C$: duplicate-pair rate.}
\label{tab:motivation-redundancy}
\footnotesize
\setlength{\tabcolsep}{4pt}
\begin{tabular}{@{}lrrrrrr@{}}
\toprule
Compiler & $B$ & $U$ & $B/U$ & ${}^2\!D$ & $R_w$ & $C$ \\
\midrule
TVM              & 127 & 4 &  31.8 & 2.12 &  60.0 & 46.8\% \\
PyTorch Inductor &  49 & 3 &  16.3 & 1.46 &  33.5 & 67.7\% \\
ONNX Runtime     & 276 & 2 & 138.0 & 1.03 & 268.1 & 97.1\% \\
\bottomrule
\end{tabular}
\end{table}
Table~\ref{tab:motivation-redundancy} shows that redundancy is severe on all
three DL compilers. On TVM, 127 reports correspond to only 4 defects
($R_w = 60.0$). PyTorch Inductor produces fewer reports, 49 for 3 defects
($R_w = 33.5$), but they are more concentrated: two randomly chosen reports
share a root cause with probability 67.7\%, compared with 46.8\% on TVM.
ONNX Runtime is the most extreme, with 276 reports for 2 defects
($R_w = 268.1$, $C = 97.1\%$). On every compiler, a single defect dominates,
accounting for 64.6\%, 81.6\%, and 98.6\% of the reports on TVM, PyTorch
Inductor, and ONNX Runtime, respectively (Table~\ref{tab:motivation-bugs}).
As a result, the effective number of defects ${}^2\!D$ is far below $U$,
ranging from 1.03 to 2.12. Each redundant report costs a compilation, an
execution, and developer triage without revealing a new defect. The skew,
however, is also an opportunity: suppressing the triggers of a few frequently
triggered defects would remove most redundant reports.

\subsection{Existing Techniques Do Not Prevent Rediscovery}
\label{sec:motivation-existing}

This redundancy of defect reports persists even though NeuRI already diversifies program generation by
inferring operator rules and combining operators randomly. Diversifying the
programs a fuzzer generates does not diversify the faults they trigger,
because many structurally different programs can reach the same faulty code
path. Swarm testing~\cite{swarm}, which randomly omits features from program
generation, shares this limitation. Coverage feedback~\cite{liu2022coverage,yang2025may}
does not solve the problem either, since a program that rediscovers a known
defect may still reach new code elsewhere. Techniques that learn from past
bugs, such as HiCOND~\cite{hicond}, steer generation toward bug-prone
configurations rather than away from known defects. Finally, deduplication
techniques rank, cluster, or bisect bug-triggering programs after program
generation~\cite{chen2013taming,yang2023d3,zhou2025bisection}. They reduce
triage effort, but only after the fuzzer has generated and executed the
duplicates. Avoiding rediscovery requires a capability that none of these techniques
provide: \emph{a reusable description of the conditions that trigger a known
defect, which the fuzzer can check during generation.}

\subsection{DL Compiler Graphs Make Trigger Conditions Describable}
\label{sec:pattern-motivation}

Such descriptions are hard to obtain for traditional compiler fuzzers.
AFL~\cite{afl} represents inputs as byte strings. Csmith~\cite{csmith}
generates C programs whose compiler-relevant behavior depends on interacting
features such as aliasing, pointers, and control flow. In both settings,
syntactic similarity is a poor proxy for compiler behavior: different programs
may share the same semantic trigger, while similar programs may exercise
different compiler paths. This difficulty helps explain why redundancy of defect reports in
compiler fuzzing is usually handled after program generation.

DL compilers offer a more suitable representation. Their inputs are
computation graphs whose nodes are tensor operations and whose edges are data
flow, annotated with properties such as shape and data type(dtype). This structure
provides a compact vocabulary for describing classes of inputs that exercise
similar compiler behavior. Compilers themselves reason in this vocabulary.
PyTorch Inductor, for example, uses pattern matching over its FX graphs to
find subgraphs it can rewrite, such as folding a batch norm into the
preceding convolution~\cite{torchinductor-patternmatcher}, and TVM fuses chains of operators into a single kernel based on
each operator's category
based on operator categories and graph structure~\cite{tvm-fuseops}. Graph
generators such as NNSmith~\cite{liu2023nnsmith} already maintain these
properties during synthesis, so patterns are cheap to check before
a program is compiled.

Topology alone, however, is not enough for reproducing a defect. Graphs with the same
\emph{operator $\rightarrow$ reshape $\rightarrow$ operator} topology may
trigger different transformations depending on whether the first operator is
elementwise or a reduction. Tensor dtypes, shapes, and operator attributes can
likewise change compiler behavior. We therefore enrich graph topology with
semantic properties available during generation: operator category, tensor
dtype and shape, selected operator attributes, and graph-level context such as
constant inputs. The resulting \emph{semantic graph patterns} capture
recurring bug-triggering conditions without requiring exact program matches.

\subsection{Conclusive Summary}
\label{sec:motivation-summary}

Three observations motivate our approach. First, DL compiler fuzzers
rediscover known defects at high rates, and this redundancy is concentrated:
a single defect accounts for 64.6\%--98.6\% of each compiler's crash reports.
Second, existing diversification and deduplication techniques do not prevent reporting known defects, because none of the techniques uses the defects already found to steer
generation away from program generations. Third, DL compiler graphs expose semantic
properties that make bug-triggering conditions describable and cheap to check
during program generation. Building on these observations, we present \toolName{}, a
pattern-guided fuzzer for DL compilers that suppresses reports of known defects. Its key idea is to turn confirmed bug-triggering patterns into
\emph{negative guidance}: when a generated node completes a known pattern,
\toolName{} locally regenerates it with a different semantic configuration,
avoiding the known trigger before compilation and execution.

\section{Approach}

\begin{figure}
    \centering
    \includegraphics[width=0.9\linewidth]{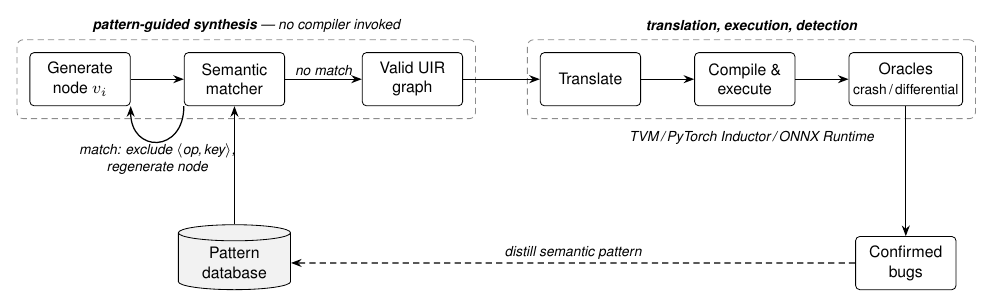}
    \caption{The \toolName{} framework.}
    \label{fig:framework}
\end{figure}

\subsection{Overview}
\label{sec:approach-overview}

To address the redundancy problem in triggering programs, we design and implement
a pattern-guided fuzzer for DL compilers, \toolName{}. Figure~\ref{fig:framework}
illustrates the overall architecture. Test programs are synthesized in a
compiler-independent unified intermediate representation (UIR), whose typed
operator signatures ensure that every generated graph is shape-valid by
construction. During synthesis, each newly created node is checked against a
database of \emph{semantic graph patterns} distilled from previously confirmed
bugs. When a match is detected, the generator regenerates the corresponding node
instead of discarding the entire graph. The resulting programs are then
translated to the target compilers and checked by a differential oracle and a crash oracle. Newly confirmed
faults are subsequently distilled into semantic graph patterns and added to the
pattern database, forming a closed feedback loop.

% \textcolor{blue}{The generated programs are checked by two oracles: a differential oracle that compares the outputs of the compiled and uncompiled executions of the same program, and a crash oracle that detects crashes, hangs, assertion failures, and timeouts during compilation or execution.}. 

The key design point is \emph{when} this check is performed. Each pattern is a
predicate over properties that the generator has already computed during
synthesis. Consequently, pattern matching requires only a lightweight lookup,
allowing redundant structures to be suppressed before translation, compilation,
and execution incur their associated costs. Rather than using known bug-triggering
structures as generation targets, \toolName{} uses these structures as \emph{negative
guidance}: once a newly generated structure resembles a previously observed trigger, 
the generator steers away from it while preserving the rest of the program. 
This enables \toolName{} to suppress reports of known defects without
discarding otherwise useful test graphs.

\subsection{Unified Intermediate Representation (UIR)}
\label{sec:uir}

A UIR program is a directed acyclic graph (DAG) of typed operators and tensors. Each operator defines a \emph{tensor signature}—specifying input/output types, attributes, and forward shape-derivation rules. By connecting only signature-compatible operators, synthesized graphs are shape-valid \emph{by construction} without expensive solver invocations (different from NNSmith~\cite{liu2023nnsmith}). This forward design makes generation lightweight and exposes node types, shapes, and attributes at node creation time, enabling \emph{incremental} pattern matching during graph construction.

The primary trade-off is expressiveness: UIR cannot naturally express backward shape constraints (e.g., satisfying a required output shape like $(1,1024)$ by solving for a compatible input like $(1,32,32)$). Although this restriction excludes certain shape-constrained graphs, it maintains high generation speed and immediate pattern-matching visibility.

UIR is backend-independent, enabling unified pattern definitions be lowered directly to TVM Relax, \texttt{torch.compile}, or ONNX. To ensure cross-compiler portability, UIR incorporates a core vocabulary of 86 operators across 18 functional categories (e.g., arithmetic, activations, shape transformations, convolutions) with semantics derived from TVM Relax, ONNX, and PyTorch APIs.

Each UIR node samples attributes and tensors within valid domains to guarantee execution stability.
\emph{Attributes} are constrained by input properties (e.g., reduction axes bounded by rank, convolution parameters capped by spatial extent).
\emph{Tensors} use number of dimensions in $[1, 4]$, dimension extents capped to $32$ with element budget limits to prevent out-of-memory errors, and standard data types (primarily \texttt{float32}).
\emph{Input values} are initialized within bounded intervals (e.g., $[-1, 1]$) with guards against numerical edge cases (e.g., zero-division or non-positive log inputs) to eliminate false-positive crashes.

\subsection{Semantic Graph Patterns}
\label{sec:pattern-def}

A pattern is a partial description of a bug-triggering UIR subgraph with four
parts. \emph{(i) Nodes:} a sequence of operator specifications, each fixing the
operator type, its input and output arity, and optionally exact values for
selected attributes; attributes left unlisted are unconstrained. \emph{(ii)
Values:} constraints on the tensors of the subgraph---rank, per-dimension extents
(exact or ranged), data type, and optionally relations across dimensions, such as
a bound on a product of dimensions. \emph{(iii) Graph constraints:} optional
conditions on the graph outside the matched subgraph, in two forms: the presence
of a required operator elsewhere in the graph, and a restriction on how the
subgraph's inputs are produced. \emph{(iv) Flow constraints:} optional data-flow
relations asserting that one node's output is consumed at a specified input
position of another node. 

% Because these relate node outputs to node inputs, a
% value that is a graph input rather than a node output cannot be the source of such
% a relation. 
% Figure~\ref{fig:pattern-parts} annotates the four parts on a concrete pattern of \ptIssue{189803}.

% Requires in preamble: \usepackage{tikz} and \usetikzlibrary{arrows.meta,calc}
% Natural width 240.8pt: fits one ACM column at full size, so no \resizebox.
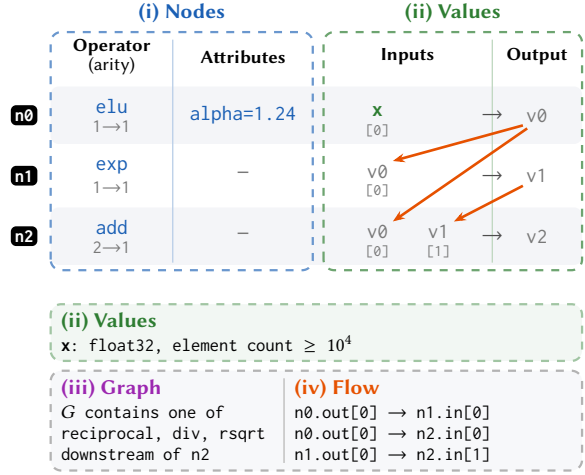
\begin{figure}[t]
\centering
\definecolor{nodesblue}{RGB}{26,95,180}
\definecolor{valgreen}{RGB}{46,125,50}
\definecolor{graphpurple}{RGB}{156,39,176}
\definecolor{floworange}{RGB}{230,81,0}
\definecolor{othergray}{RGB}{120,120,120}
\definecolor{zebragray}{RGB}{244,245,248}
\begin{tikzpicture}[
  x=0.95cm, y=1cm,
  badge/.style={rounded corners=2pt, fill=black, text=white,
                font=\bfseries\ttfamily\scriptsize, inner sep=1.5pt},
  grpbox/.style={dashed, line width=0.85pt, rounded corners=4pt},
  hdr/.style={font=\bfseries\sffamily\scriptsize, align=center},
  tag/.style={font=\bfseries\sffamily\footnotesize, align=center},
  code/.style={font=\ttfamily\footnotesize, align=center, inner sep=0.5pt},
  sub/.style={font=\ttfamily\scriptsize, text=othergray, align=center, inner sep=0.5pt},
  posn/.style={font=\ttfamily\tiny, text=othergray, inner sep=0.3pt},
  flow/.style={-{Stealth[length=1.6mm]}, line width=1.0pt, draw=floworange},
  para/.style={font=\footnotesize, anchor=north west, align=left, inner sep=1pt,
               execute at begin node=\frenchspacing},
]
\fill[zebragray, rounded corners=3pt] (0.75, 0.42) rectangle (8.15, 1.18);
\fill[zebragray, rounded corners=3pt] (0.75,-1.18) rectangle (8.15,-0.42);

\draw[grpbox, draw=nodesblue!65] (0.75,-1.25) rectangle (4.40, 1.90);
\draw[grpbox, draw=valgreen!70]  (4.55,-1.25) rectangle (8.15, 1.90);
\draw[draw=nodesblue!30, line width=0.5pt] (2.45,-1.15) -- (2.45, 1.80);
\draw[draw=valgreen!30,  line width=0.5pt] (6.90,-1.15) -- (6.90, 1.80);

\node[tag, text=nodesblue] at (2.575, 2.15) {(i) Nodes};
\node[tag, text=valgreen]  at (6.35,  2.15) {(ii) Values};

\node[hdr] at (1.60, 1.58) {Operator\\[-1pt]{\mdseries\scriptsize(arity)}};
\node[hdr] at (3.43,  1.58) {Attributes};
\node[hdr] at (5.72,  1.58) {Inputs};
\node[hdr] at (7.52,  1.58) {Output};

\node[badge] at (0.38, 0.8)  {n0};
\node[badge] at (0.38, 0.0)  {n1};
\node[badge] at (0.38,-0.8)  {n2};

% n0: ceil -- produces -0.0 from inputs in (-1,0)
\node[code, text=nodesblue] at (1.60, 0.92) {elu};
\node[sub]                  at (1.60, 0.65) {1$\to$1};
\node[code, text=nodesblue] at (3.43, 0.8)  {alpha=1.24};
\node[code, text=valgreen, font=\bfseries\ttfamily\footnotesize] at (5.30, 0.86) {x};
\node[posn]                 at (5.30, 0.60) {[0]};
\node[code, text=othergray] (o0) at (7.52, 0.8) {v0};

% n1: zeros_like
\node[code, text=nodesblue] at (1.60, 0.12) {exp};
\node[sub]                  at (1.60,-0.15) {1$\to$1};
\node[code, text=othergray] at (3.43, 0.0)  {---};
\node[code, text=othergray] (i1) at (5.30, 0.06) {v0};
\node[posn]                 at (5.30,-0.20) {[0]};
\node[code, text=othergray] (o1) at (7.52, 0.0) {v1};

% n2: add -- Inductor drops the IEEE-754 sign of zero here
\node[code, text=nodesblue] at (1.60,-0.68) {add};
\node[sub]                  at (1.60,-0.95) {2$\to$1};
\node[code, text=othergray] at (3.43,-0.8)  {---};
\node[code, text=othergray] (i2a) at (5.30,-0.74) {v0};
\node[posn]                 at (5.30,-1.00) {[0]};
\node[code, text=othergray] (i2b) at (6.15,-0.74) {v1};
\node[posn]                 at (6.15,-1.00) {[1]};
\node[code, text=othergray] (o2) at (7.52,-0.8) {v2};

\foreach \y in {0.8, 0.0, -0.8}
  \node[font=\footnotesize, text=black!70] at (6.90, \y) {$\rightarrow$};

% (iv) flow: v0 fans out to n1 and n2 -- the diamond
\draw[flow] ($(o0.south west)+(-0.02,-0.02)$) -- ($(i1.north east)+(0.04,0.02)$);
\draw[flow] ($(o0.south west)+(0.04,-0.06)$)  -- ($(i2a.north east)+(0.04,0.02)$);
\draw[flow] ($(o1.south west)+(-0.02,-0.02)$) -- ($(i2b.north east)+(0.04,0.02)$);

% (ii) values
\fill[valgreen!6, rounded corners=4pt] (0.75,-2.45) rectangle (8.15,-1.70);
\draw[grpbox, draw=valgreen!45] (0.75,-2.45) rectangle (8.15,-1.70);
\node[para] at (0.85,-1.76)
  {\textbf{\sffamily\color{valgreen}(ii) Values}\\[-1pt]
   {\ttfamily\scriptsize \textbf{x}: float32, element count $\ge$ $10^4$}};

% (iii) graph + (iv) flow
\fill[zebragray!70, rounded corners=4pt] (0.75,-3.90) rectangle (8.15,-2.58);
\draw[grpbox, draw=black!30] (0.75,-3.90) rectangle (8.15,-2.58);
\draw[draw=black!20, line width=0.7pt] (4.00,-3.80) -- (4.00,-2.68);
\node[para] at (0.85,-2.65)
  {\textbf{\sffamily\color{graphpurple}(iii) Graph}\\[-1pt]
   {\ttfamily\scriptsize $G$ contains one of}\\[-2pt]
   {\ttfamily\scriptsize reciprocal, div, rsqrt}\\[-2pt]
   {\ttfamily\scriptsize downstream of n2}};
\node[para] at (4.10,-2.65)
  {\textbf{\sffamily\color{floworange}(iv) Flow}\\[-1pt]
   {\ttfamily\scriptsize n0.out[0] $\to$ n1.in[0]}\\[-2pt]
   {\ttfamily\scriptsize n0.out[0] $\to$ n2.in[0]}\\[-2pt]
   {\ttfamily\scriptsize n1.out[0] $\to$ n2.in[1]}};
\end{tikzpicture}
\caption{The semantic graph pattern distilled from \ptIssue{189803}. }
\label{fig:pattern-parts}
\end{figure}

Figure~\ref{fig:pattern-parts} shows a pattern from a bug we found, distilled from
\ptIssue{189803}. For negative inputs the quantity \texttt{elu(x)+exp(elu(x))}
passes through zero, and \texttt{torch.compile} loses the cancellation, differing
from eager execution by up to $108$ in absolute value. Each part carries a condition the fault requires: \emph{(i)} fixes three operator
types with their arities and pins \texttt{alpha} on \texttt{elu}, leaving every
other attribute free; the value matters, because \texttt{elu} is bounded below by
$-\texttt{alpha}$ and the sum reaches zero only for arguments near $-0.568$.
\emph{(ii)} requires \texttt{float32} and at least $10^4$ input elements:
the cancellation occurs only for input values in a narrow interval, and
because element values are drawn at random, a small tensor is unlikely to
contain any value in that interval. 
\emph{(iii)} requires a reciprocal-like operator somewhere downstream. On its
own, the lost cancellation changes the result by only a small absolute
amount; a downstream reciprocal
divides by this near-zero value and amplifies the difference into a
reportable failure. \emph{(iv)} states that \texttt{v0} is consumed by both \texttt{exp} and
\texttt{add}, so the pattern is a diamond rather than a chain.

The pattern generalizes the reproducer rather than transcribing it: the shape of
the input, its element values, and the choice of downstream operator are left
open. It also illustrates the incompleteness discussed above, since a graph whose
inputs never drive the sum near zero matches the pattern without triggering the
fault. A pattern fires only when every node specification and every value, graph, and
flow constraint is satisfied; partial matches only track progress and never
trigger regeneration. 

\paragraph{Semantic key.}
The \emph{semantic key} of a node $v$ with respect to a pattern $p$,
$\textsc{Key}_p(v)$, is the projection of $v$ onto the features that $p$
constrains at $v$'s position (fixed attributes, dtype, rank, constrained
extents, and producer operator kinds). Ranged constraints contribute the cell
they define rather than the exact value (e.g., ``element count $\ge 10^4$''
contributes one bit), and unconstrained features are dropped. Two candidates
with the same operator and key are thus indistinguishable to $p$, so excluding
one excludes only the matched configuration. The key is computed from
node-local information, requiring no solver call or graph traversal.

\subsection{Distilling and Tightening Patterns}
\label{sec:distill}

Patterns are written manually. The pattern database starts empty; when a new
bug-triggering program is confirmed, we first reduce it to a minimal
reproducer and then generalize the reproducer into a semantic pattern. This requires identifying which features of the reproducer are essential to
the fault and which can safely remain unconstrained; the features retained in
the pattern are exactly those that form its semantic key. The decision is guided
by the compiler's failure behavior---for example, whether the fault depends
on an operator or fusion category, a particular tensor extent, a data type,
or a dispatch-relevant attribute. This is the only step in \toolName{} that
requires manual analysis. Writing one pattern took a median of
$30$ minutes by two authors with general fuzzing knowledge. Writing a pattern is a one-time cost for each fault: once written, the
pattern is reused in every later run. Without it, every run would again
produce duplicate reports of that fault for someone to triage.

An overly general pattern may match programs that do not trigger the
corresponding bug, causing \toolName{} to regenerate valid candidates
unnecessarily. We therefore validate each newly distilled pattern against a
held-out corpus disjoint from both the programs used for distillation and
those inspected during the corresponding fuzzing run. If a pattern matches
a program that compiles and executes correctly, we add a value or graph
constraint that separates the false match from the known trigger and
revalidate the pattern. We repeat this process until no false matches remain
or the pattern reaches a limit of 10 tightening rounds. A pattern that still
produces false matches after 10 rounds is discarded.

\subsection{Pattern-Guided Synthesis}
\label{sec:synthesis}

The key design of \toolName{} is to use previously observed bug-triggering
structures as \emph{negative guidance during synthesis}, rather than filtering
completed programs after generation. Algorithm~\ref{alg:gen} integrates this
guidance directly into graph construction. The
generator proposes a candidate node and incrementally checks whether it
completes a known pattern. A candidate that does not complete a pattern is
generated immediately. When a known trigger is recognized, the candidate is
discarded and an alternative semantic configuration is generated at the same
position. The already constructed graph prefix is preserved.

\begin{algorithm}[!t]
\caption{Pattern-Guided Graph Generation}
\label{alg:gen}
\begin{algorithmic}[1]
\Require graph size $n$, pattern database $\mathcal{P}$, retry bound $R$
\Ensure a shape-valid UIR graph $G$
\State $G \gets \emptyset$; \quad $\Gamma \gets \emptyset$ \Comment{live prefixes}
\For{$i \gets 1$ \textbf{to} $n$}
  \State $B \gets \emptyset$ \Comment{keys excluded here} \label{lin:reset}
  \For{$r \gets 1$ \textbf{to} $R$}
    \State $v \gets \Call{GenerateNode}{G, B}$ \label{lin:gen}
    \State $(p, \Gamma') \gets \Call{PatternMatch}{v, \mathcal{P}, \Gamma}$ \label{lin:match}
    \If{$p = \textbf{null}$ \textbf{or} $r = R$}
      \State $G \gets G \cup \{v\}$; \quad $\Gamma \gets \Gamma'$ \label{lin:commit}
      \State \textbf{break}
    \EndIf
    \State $B \gets B \cup \{\langle p, \textsc{Op}(v), \textsc{Key}_p(v)\rangle\}$ \label{lin:exclude}
  \EndFor
\EndFor
\State \Return $G$
\end{algorithmic}
\end{algorithm}

Algorithm~\ref{alg:gen} builds a graph node at a time. At each of the $n$
positions, the generator proposes a candidate node $v$ that is signature-valid
with respect to the current graph $G$ and whose semantic key is not in the
exclusion set $B$ (Line 5). The matcher then checks whether
$v$ completes a known bug-triggering pattern, using the live prefixes $\Gamma$
to track patterns partially matched by earlier nodes
(Line 6). The matcher does not modify $\Gamma$; instead,
it returns the prefix state $\Gamma'$ that would result if $v$ were accepted. If $v$ completes no pattern, it is added to $G$ and $\Gamma'$ becomes the current
prefix state (Line 8). If it completes a known pattern $p$, $v$ is discarded and the triple
$\langle p, \textsc{Op}(v), \textsc{Key}_p(v)\rangle$ is added to $B$
(Line 11), so the next attempt at this position cannot propose
a candidate that agrees with $v$ on operator type and on $p$'s key.

Three choices keep the guidance from narrowing the search more than necessary.
First, exclusion is by semantic key rather than operator type: a pattern
implicates a particular configuration of an operator, not the operator itself.
Thus, a matched \textsc{conv2d} can be retried with a different shape or
attribute rather than being removed from the search. Second, the exclusion set
is reset at every position (Line 3), so it affects only
the current node and never constrains later positions. Third, retries are
bounded by $R$: if every attempt at a position completes a known pattern, the
last candidate is accepted anyway (Line 8). Generation
therefore never stalls at a position because of pattern avoidance. The matching overhead for each generated
node is $O(|\Gamma|+|\mathcal{P}|)$, where $|\Gamma|$ is the number of live
prefixes; In practice, $|\Gamma|$ stays small because a
prefix starts only when a node matches the first node of some pattern.

\subsection{Translation and Bug Detection}\label{sec:system}

\toolName{} translates each synthesized UIR program into the representation
required by the target compiler and executes it through the corresponding
compilation pipeline. For TVM, the UIR graph is lowered to a Relax program and
compiled with the CUDA backend. For PyTorch, the graph is expressed as a
module and compiled with \texttt{torch.compile}. For ONNX Runtime, the graph is
serialized as an ONNX model and passed through its optimization and execution
pipeline.

% \paragraph{Oracles.}
% \toolName{} uses two types of oracles. A \emph{compilation oracle} reports
% aborts, assertion failures, or sanitizer violations encountered during
% compilation. Because the UIR generator guarantees shape validity before
% translation, such failures indicate faults in the compiler rather than invalid
% test inputs. A \emph{differential oracle} compares the behavior of compiled
% programs against an independent reference. For PyTorch, each program is
% executed both eagerly and through \texttt{torch.compile}; a mismatch in the
% resulting tensors indicates a potential miscompilation. For ONNX Runtime, the
% model is executed with and without graph optimization, and a difference in
% outputs indicates a potential optimizer defect. Compilation or execution
% failures are recorded as crashes. Every reported bug is independently
% reproduced from the complete source before being counted
% (Section~\ref{sec:rq3}).

% \paragraph{Feedback loop.}
Confirmed bugs are analyzed and distilled into semantic graph patterns, which
are added to the pattern database for subsequent runs. Thus, bug detection
feeds directly back into synthesis: newly confirmed triggering structures
become negative guidance that helps \toolName{} avoid rediscovering the same
faults while exploring new program configurations.

\section{Experiment Setup}
\label{sec:setup}

\subsection{Research Questions}
\label{sec:rqs}

\begin{itemize}[leftmargin=*]
\item \textbf{RQ1: Redundancy reduction} (Section~\ref{sec:rq1}).
How effectively does semantic pattern guidance reduce the rate of
rediscovering previously encountered defects?

\item \textbf{RQ2: Cost of pattern guidance} (Section~\ref{sec:rq2}).
What trade-offs does \toolName{} incur in branch coverage and generation
throughput when reducing redundancy, compared with the state-of-the-art
fuzzer and an unguided variant of \toolName{}?

\item \textbf{RQ3: Bug-finding effectiveness} (Section~\ref{sec:rq3}).
Can \toolName{} discover previously unknown bugs in real-world DL compilers?

\end{itemize}

\subsection{Experimental Setup}
\label{sec:env}

\textbf{Systems under test (SUT):} We evaluate three widely used DL compilers with distinct graph
transformation and code-generation pipelines. \emph{TVM}~\cite{tvm} 0.25.0 
compiles models through its Relax graph IR and TIR before generating target
code. \emph{PyTorch Inductor} ~\cite{pytorch} 2.13.0, the backend of
\texttt{torch.compile}, lowers captured FX graphs through ATen to Triton and
C++ kernels. \emph{ONNX Runtime}~\cite{onnxruntime} v1.27.0 optimizes ONNX
computation graphs, partitions them across execution providers, and executes
the resulting subgraphs using provider-specific kernels.

\noindent
\textbf{Test oracles:}
Following prior work~\cite{liu2023neuri,liu2023nnsmith}, we use two test
oracles. \emph{Result inconsistency} indicates a bug when the compiled model
produces an output that differs from eager execution by more than an absolute
error of $10^{-3}$ and a relative error of 1\%. \emph{Runtime error} indicates
a bug when compilation or execution fails unexpectedly. 

% Exceptions raised
% during eager execution are treated as invalid test cases rather than bugs.
\noindent
\textbf{Branch coverage:}
We measure cumulative branch coverage 
of generated programs~\cite{zhu1997coverage}. Coverage is collected
from instrumented builds of each SUT, and the same builds are used for both
coverage measurement and bug discovery. For TVM, we use commit
\texttt{b3e249b7} (\href{https://github.com/apache/tvm/releases/tag/v0.25.0.post1}{v0.25.0.post1}, 2026-06-24) and build in RelWithDebInfo
mode with \texttt{-fprofile-arcs -ftest-coverage}, CUDA support, and LLVM~15
for host code generation. We instrument \texttt{src/relax},
\texttt{src/tir}, and \texttt{src/target}, which cover the components exercised
by the Relax-to-CUDA compilation path. For PyTorch, we use commit
\texttt{af333cf2} (\href{https://github.com/pytorch/pytorch/releases/tag/v2.13.0-rc9}{v2.13.0-rc9}, 2026-06-23) with
\texttt{USE\_CPP\_CODE\_COVERAGE=1}. Following NeuRI~\cite{liu2023neuri}, we
instrument non-kernel files under \texttt{torch/csrc}, \texttt{aten/}, and
\texttt{c10/}, whose code collectively implements the lowering path exercised
by PyTorch Inductor. For ONNX Runtime, we use commit \texttt{8f0278c}
(\href{https://github.com/microsoft/onnxruntime/releases/tag/v1.27.0}{v1.27.0}, 2026-06-12) built with the same coverage instrumentation
(\texttt{-fprofile-arcs -ftest-coverage}), and measure coverage over its core
(\texttt{onnxruntime/core}), which implements the graph optimization,
execution-provider partitioning, and kernel-dispatch path exercised by the
generated ONNX graphs.

\noindent
\textbf{Baselines:}
We compare against NeuRI~\cite{liu2023neuri}, a generation-based DL
compiler fuzzer that has been evaluated against both generation- and
mutation-based fuzzers. We also evaluate an ablated version of \toolName{},
denoted \toolName{}$^-$, with pattern-guided deduplication disabled.
\toolName{}$^-$ uses the same program synthesis pipeline and seed sequences as
\toolName{} and differs only in whether the pattern check is applied. This
controlled comparison isolates the effect of pattern-guided deduplication
from other components of the generator.

\noindent
\textbf{Configuration:}
All experiments run on a machine with an AMD Ryzen 9 9950X
CPU (32 threads, 16 cores), 64\,GB of memory, and an NVIDIA RTX 3080 Ti GPU (12\,GB),
running Ubuntu 24.04.

\section{Results}
\label{sec:results}

\subsection{RQ1: Redundancy Reduction}
\label{sec:rq1}

We investigate whether pattern guidance reduces repeated crash reports
for known defects. On TVM (CUDA), we compare four configurations that
differ only in the active patterns: \textsf{none} (\toolName$^-$),
\textsf{only A} (two patterns for \tvmIssue{20036}, a batch-norm CUDA
code-generation failure), \textsf{only B} (four patterns for
\tvmIssue{20047}, a \texttt{conv2d} GPU scheduling assertion), and
\textsf{A$\cup$B} (all six patterns). Each configuration generates
$N=10{,}000$ programs from the same seed sequence, with one run per configuration. Because generation
diverges after the first rejection, this is an end-to-end comparison
under a fixed generation budget, not a comparison of paired program
outcomes. On PyTorch Inductor, we compare \textsf{none} with the full
pattern set. We analyze deterministic
crashes only, assigning root causes as described in
Section~\ref{sec:motivation}. Since the patterns were distilled from
reproducers of the defects they target, this RQ evaluates suppression
of known defects rather than generalization to unseen defects.

We report $B$ (crash reports), $U$ (distinct defects), and $B-U$, the reports that repeat a defect already seen in the run, together with other metrics: ${}^2\!D$, $R_w$, and $C$ (see Section 2). $B$ and $B-U$ measure the \emph{volume} of reports a developer must triage. ${}^2\!D$, $R_w$, and $C$ measure how reports concentrate over the defects that \emph{survive}. When $U=1$ these reports become trivial ($R_w=B$, ${}^2\!D=1$, $C=100\%$), so an increase in $C$ after suppression reflects fewer surviving defects, not more redundancy. We therefore compare configurations on $B$ and $B-U$ and use the other ${}^2\!D$, $R_w$, and $C$ metrics to describe each distribution. We also report the largest portion of defects report share, which stays above 90\% in every
configuration (TVM: 98.8\% to 96.0\%; PyTorch: 99.5\% to 91.9\%). So guidance
reduces the volume of duplicate reports but not their dominance.

\begin{table}[t]
\centering
\caption{Crash reports with and without patterns over 10{,}000 programs per configuration, one run each. Per-defect counts $b_d$ follow the order \tvmIssue{20036}/\tvmIssue{20047}/\tvmIssue{20048} for TVM and \ptIssue{198613}/\ptIssue{193581}/\ptIssue{195963} for PyTorch. }
\label{tab:rq1-ablation-breakdown}
\small
\setlength{\tabcolsep}{3.5pt}
\begin{tabular}{llrrrrrrrr}
\toprule
\textbf{Compiler} & \textbf{Config} ($b_d$) & $B$ & $U$ & $B{-}U$ & $\Delta B$ & ${}^2\!D$ & $R_w$ & $C$ \\
\midrule
TVM (CUDA)
  & \textsf{none} (216/36/2)    & 254 & 3 & 251 & ---       & 1.35 & 188.8 & 74.2\% \\
  & \textsf{only A} (0/30/0)    & 30  & 1 & 29  & $-88.2\%$ & 1.00 & 30.0  & 100\%  \\
  & \textsf{only B} (209/26/0)  & 235 & 2 & 233 & $-7.5\%$  & 1.25 & 188.8 & 80.2\% \\
  & \textsf{A$\cup$B} (0/25/0)  & 25  & 1 & 24  & $-90.2\%$ & 1.00 & 25.0  & 100\%  \\
\midrule
PyTorch
  & \textsf{none} (563/27/4)    & 594 & 3 & 591 & ---       & 1.11 & 534.9 & 90.0\% \\
  & full (1/35/1)               & 37  & 3 & 34  & $-93.8\%$ & 1.12 & 33.2  & 89.3\% \\
\bottomrule
\end{tabular}
\end{table}

Without patterns, the 10{,}000 TVM programs produce $B=254$ crashes
from $U=3$ defects. The distribution is highly skewed:
\tvmIssue{20036} accounts for 85.0\% (216/254) of reports,
\tvmIssue{20047} for 14.2\% (36/254), and \tvmIssue{20048} for
0.8\% (2/254). The report-weighted redundancy is $R_w=188.8$, the effective
number of defects is ${}^2\!D=1.35$ (out of three), and two randomly
selected reports share a root cause with probability $C=74.2\%$.
Overall, 251 of the 254 reports repeat a defect already observed. With \textsf{only A} pattern guidance, all 216 reports of \tvmIssue{20036} disappear.
The total defect reports falls by 88.2\%, while $B-U$ falls by 88.4\% (251 to 29).
By contrast, \textsf{only B} leaves the dominant defect largely
unchanged: $B=235$, with $R_w=188.8$ and $C=80.2\%$.
The full set yields $B=25$ ($-90.2\%$) and $B-U=24$ ($-90.4\%$). With all four patterns for \tvmIssue{20047} active, 25 of its 36 reports remain,
and \textsf{only A}, which already lowers defects reports to 30.
We therefore do not attribute the gap between 88.2\% and 90.2\% to group B patterns.
Thus, most of the reduction comes from suppressing the dominant
defect. In this skewed distribution, addressing the head defect
accounts for most of the reduction in report volume. The reduction in $U$ from three to one is expected for
\tvmIssue{20036}, but the absence of \tvmIssue{20048} in guided runs
deserves caution. \tvmIssue{20048} appears only twice in the baseline, so its absence from a
single guided run is unremarkable: under the rate $p=2/10{,}000$, a
10{,}000-program run can hit this defect with probability about 13.5\%. This explains
this defect's absence in \textsf{only A}, where no pattern targets its structure. In
\textsf{only B} and \textsf{A$\cup$B}, the absence is also consistent with
chance, but \tvmIssue{20048} shares a \texttt{conv2d} scheduling subgraph with
\tvmIssue{20047}, so collateral suppression by group B is plausible. Because
$p$ rests on two observations, the data neither establish nor rule out
over-suppression.

On PyTorch Inductor, 563 of 594 baseline crashes (94.8\%) come from
\ptIssue{198613}, an AVX512 boolean-vectorization cast failure
($R_w=534.9$, $C=90.0\%$). With the full pattern set, this defect
appears once, and $B$ falls from 594 to 37 ($-93.8\%$), while
$B-U$ falls from 591 to 34 ($-94.2\%$). All three defects remain
observed. The two untargeted defects change from 27 and 4 reports
to 35 and 1, respectively. Thus, suppressing the dominant defect
does not eliminate the other observed defects, although the change
from four reports to one is too small to interpret from a single run. After suppression, the defect with 35 reports accounts for 94.6\%
of crashes. Accordingly, concentration among the surviving defects
is nearly unchanged ($C=90.0\%\to89.3\%$ and
${}^2\!D=1.11\to1.12$), while $(B-U)/B$ changes from 99.5\% to
91.9\%. As in TVM, guidance reduces report volume, but the next
most frequent defect becomes dominant. Suppressing that defect
would require the defect's own pattern, illustrating the iterative nature
of the feedback loop. The decrease in $R_w$ from 534.9 to 33.2
primarily reflects the smaller report count and should not be
interpreted as a reduction in concentration. These results are
also based on one run per configuration.

On ONNX Runtime, both configurations produce zero crashes over
10{,}000 programs (seed 42), so we make no redundancy claim.
Zero observed crashes do not establish a zero crash rate. The
276 crashes produced by NeuRI (see Section 2)
highlight that observed outcomes depend on the generator; this
difference cannot be attributed to pattern guidance. We do not
measure redundancy among ONNX Runtime miscompilations here as its root cause is difficult to cluster.

\textbf{Answer to RQ1:}
Patterns targeting dominant known defects substantially reduce
crash-report volume: by 88.2\% on TVM with group A, by 90.2\%
with the full set (254 to 25), and by 93.8\% on PyTorch Inductor
(594 to 37). All three PyTorch defects remain observed, but
concentration among them changes little. These results are based on one run per configuration and evaluate
patterns against the defects from which they were distilled;
their effectiveness on unseen defects remains to be investigated.

\subsection{RQ2: The Cost of Pattern Guidance}
\label{sec:rq2}

\begin{figure*}[t]
    \centering
    \begin{subfigure}[b]{0.33\linewidth}
        \centering
        \includegraphics[width=\linewidth]{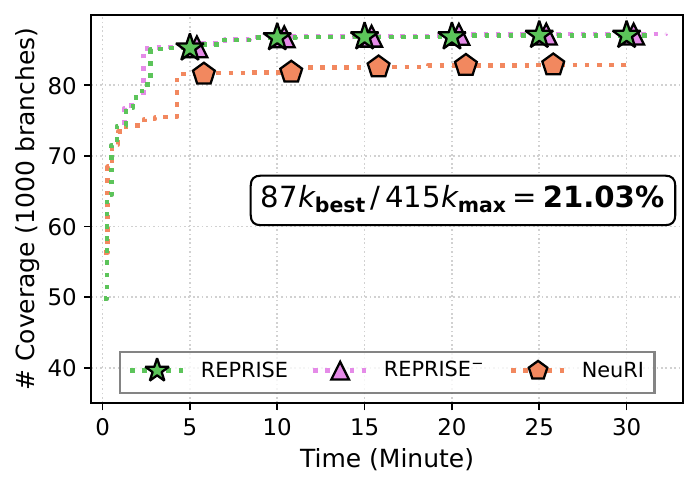}
        \caption{TVM}
        \label{fig:cov-tvm-time}
    \end{subfigure}\hfill
    \begin{subfigure}[b]{0.33\linewidth}
        \centering
        \includegraphics[width=\linewidth]{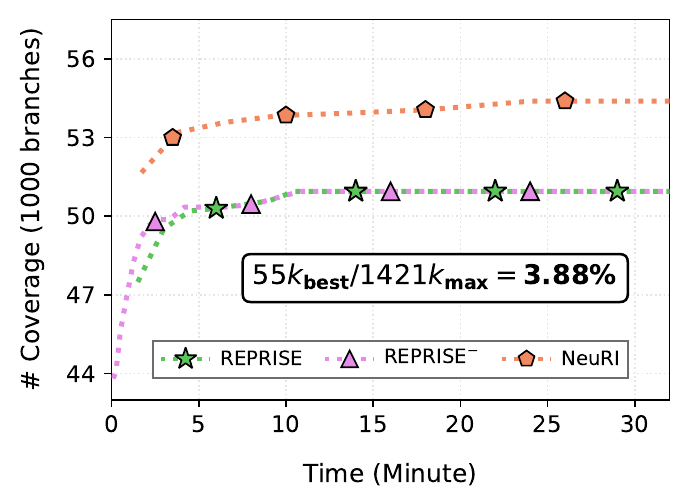}
        \caption{PyTorch Inductor}
        \label{fig:cov-pt-time}
    \end{subfigure}\hfill
    \begin{subfigure}[b]{0.33\linewidth}
        \centering
        \includegraphics[width=\linewidth]{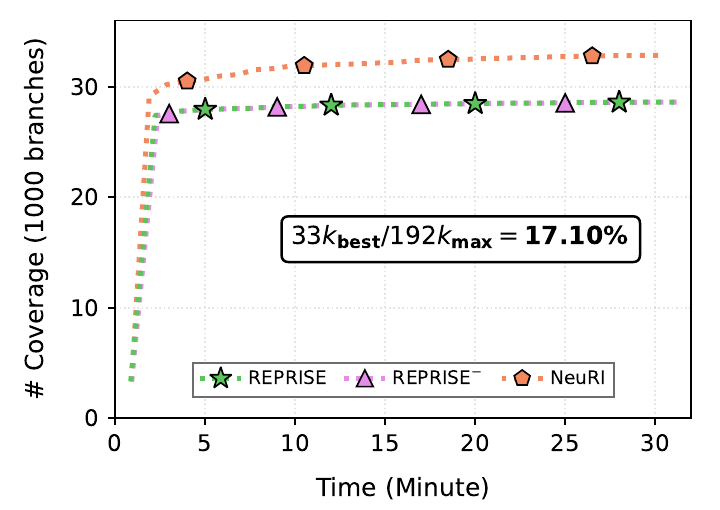}
        \caption{ONNX Runtime}
        \label{fig:cov-onnx-time}
    \end{subfigure}
    \caption{Branch coverage of \toolName{}, \toolName{}$^-$, and
    NeuRI on TVM, PyTorch Inductor, and ONNX Runtime.}
    \label{fig:coverage}
\end{figure*}

Pattern guidance rejects candidate graphs during synthesis and can therefore
reduce both generation throughput and, potentially, coverage. We first quantify
these costs through a controlled comparison with an unguided variant of
\toolName{}, and then compare \toolName{} with the state-of-the-art of DL compiler fuzzer. We use a
fixed budget of 30 minutes because, in our preliminary runs, branch coverage
on all three compiler backends had effectively plateaued by this duration, with
no further observable increase after 30 minutes. Thus, extending the budget
would add execution cost without materially changing the coverage reached by
these experiments. Under this fixed budget, we run three configurations on
three compiler platforms. The first configuration is \toolName{}. The second is
\toolName{}$^-$, which is identical to \toolName{} except that pattern-guided
suppression is disabled and the same seed sequences are used. The third is
NeuRI~\cite{liu2023neuri}. We run each configuration \nRepeats
times with distinct seed sequences and measure branch coverage and the number
of test cases executed. Figure~\ref{fig:coverage} shows coverage growth over
time, while Table~\ref{tab:throughput} reports the number of executed test
cases within the fixed budget.

\begin{table}[t]
\centering
\caption{Test cases executed within the fixed budget (30 mins).}
\label{tab:throughput}
\footnotesize
\setlength{\tabcolsep}{3pt}
\begin{tabular}{@{}lrrrr@{}}
\toprule
Tool / Metric & \shortstack[c]{TVM (CUDA)} & \shortstack[c]{PyTorch Inductor (CPU)} & \shortstack[c]{ONNX Runtime (CPU)} \\
\midrule
\toolName{}          & 1{,}540  & 4{,}280 & 30{,}135 \\
\toolName{}$^-$      & 1{,}555 & 4{,}615 & 36{,}720 \\
NeuRI                & 554     & 3{,}899 & 71{,}935 \\
\midrule
\text{Tests rejected by guidance} & 15 (1.0\%)  & 335 (7.3\%) & 6{,}585 (17.9\%) \\
\bottomrule
\end{tabular}
\end{table}

\paragraph{Guidance has little effect on coverage.}
On TVM (CUDA), \toolName{} covers 87{,}228 branches, compared with
87{,}360 for \toolName{}$^-$, a difference of only 132 branches (0.03
percentage points). On PyTorch (\texttt{torch.compile}, CPU), the two
configurations cover 50{,}954 and 50{,}950 branches, respectively, a difference
of only 4 branches. On ONNX Runtime (CPU), both configurations cover
28{,}605 branches. Across the repeated runs, these differences are small
relative to run-to-run variation and do not indicate a measurable coverage
advantage from disabling pattern guidance. Thus, under our experimental
budget, guidance incurs little observable coverage loss.

\paragraph{Guidance reduces throughput.}
Table~\ref{tab:throughput} reports the number of test cases executed within the
30-minute budget. Compared with \toolName{}$^-$, \toolName{} executes 1.0\%, 7.3\%, and 17.9\% fewer test cases on TVM (CUDA),  PyTorch
(\texttt{torch.compile}, CPU), and ONNX Runtime (CPU), respectively. The corresponding rejected
candidate counts are 15, 335, and 6585. These rejections account
directly for the throughput reduction: \toolName{} spends part of its
generation budget filtering candidates that match previously identified
bug-triggering patterns. Despite executing fewer tests, \toolName{} achieves
comparable coverage to \toolName{}$^-$ in all three compiler platforms,
indicating that the throughput reduction does not translate into a measurable
coverage loss under the same time budget. The substantially lower throughput on TVM with CUDA is primarily due to the
higher compilation cost of the CUDA pipeline. As a result, \toolName{} 
processes substantially fewer candidate programs within the fixed 30-minute
budget. 

\paragraph{Coverage remains competitive with the state-of-the-art.}
At the end of the runs, \toolName{} covers 87{,}228 of 415{,}289 branches
(21.00\%) on TVM (CUDA), compared with 82{,}889 (19.96\%) for NeuRI~\cite{liu2023neuri}. On
PyTorch (\texttt{torch.compile}, CPU), \toolName{} covers 51{,}031 of
1{,}421{,}393 branches (3.59\%), compared with 55{,}089 (3.88\%) for NeuRI.
On ONNX Runtime (CPU), \toolName{} covers 28{,}605 of 192{,}114 branches
(14.89\%), compared with 32{,}838 (17.10\%) for NeuRI. Relative to NeuRI,
\toolName{} therefore covers 5.2\% more branches on TVM, but 7.4\% and 12.9\%
fewer branches on PyTorch Inductor and ONNX Runtime, respectively. Thus, \toolName{}
achieves higher coverage on TVM, while NeuRI achieves higher coverage on PyTorch Inductor and ONNX Runtime.

The throughput comparison also varies across platforms. On TVM (CUDA),
\toolName{} executes 1{,}540 test cases within 30 minutes, compared with 554
for NeuRI, or approximately $2.8\times$ as many.
In contrast, NeuRI executes more test cases on ONNX Runtime (71{,}935 versus
30{,}135). These differences reflect the overall cost of each testing
pipeline, including test generation, compilation, and execution; we therefore
report these results as throughput differences rather than attributing them to a
specific component without additional measurements.

\textbf{Answer to RQ2.}
Pattern guidance trades generation throughput for redundancy reduction,
without materially affecting branch coverage. Compared with the unguided variant,
\toolName{} executes fewer tests because it rejects pattern-matching
candidates, yet achieves comparable branch coverage across the three
compilers. Compared with NeuRI, its coverage is higher on TVM and lower on
PyTorch Inductor and ONNX Runtime.

% 

% \begin{table}[t]
% \centering
% \caption{Test cases executed within the fixed budget (30 mins).}
% \label{tab:throughput}
% \footnotesize
% \setlength{\tabcolsep}{3pt}
% \begin{tabular}{@{}lrrrr@{}}
% \toprule
% Tool / Metric & \shortstack[c]{TVM (CUDA)} & \shortstack[c]{TVM (CPU)} & \shortstack[c]{PyTorch Inductor (CPU)} & \shortstack[c]{ONNX Runtime (CPU)} \\
% \midrule
% \toolName{}          & 1{,}540  & 12{,}913 & 4{,}280 & 30{,}135 \\
% \toolName{}$^-$      & 1{,}555  & 16{,}303 & 4{,}615 & 36{,}720 \\
% NeuRI                & 554      & 10{,}323 & 3{,}899 & 71{,}935 \\
% \midrule
% \text{Tests rejected by guidance} & 15 (1.0\%) & 3{,}390 (20.8\%) & 335 (7.3\%) & 6{,}585 (17.9\%) \\
% \bottomrule
% \end{tabular}
% \end{table}

\subsection{RQ3: Bug-Finding Effectiveness}
\label{sec:rq3}

\begin{table*}[t]
\centering
\caption{All \numBugs{} unique bugs \toolName{} reported to the developers of TVM,
PyTorch, and ONNX Runtime.
\textsc{Status}: \textsc{Fixed} = fix merged; \textsc{Not Planned} = rejected/not planned by maintainers;
\textsc{Confirmed} = bug acknowledged but not yet fixed; \textsc{Reported} = filed, no
maintainer response yet.}
\label{tab:bugs-detail}
\small
\setlength{\tabcolsep}{2.5pt}
\begin{tabular}{lccccc}
\toprule
Framework & Bug ID & \# of Patterns & Module & Category & Status \\
\midrule
    \multirow{8}{*}{TVM}
    & \tvmIssue{20015} & 11 & LLVM CodeGen & Crash & Confirmed \\
    & \tvmIssue{20036} & 2  & CUDA CodeGen & Crash & Reported \\
    & \tvmIssue{20047} & 4  & GPU Scheduling & Crash & Confirmed \\
    & \tvmIssue{20048} & 2  & GPU Scheduling & Crash & Fixed \\
    & \tvmIssue{20060} & 1  & CUDA CodeGen & Crash & Fixed \\
    & \tvmIssue{20061} & 1  & GPU Scheduling & Crash & Reported \\
    & \tvmIssue{20117} & 1  & Onnx Convert & Crash & Fixed \\
    & \tvmIssue{20118} & 1  & CUDA CodeGen & Crash & Fixed \\
    \midrule
    \multirow{11}{*}{PyTorch}
    & \ptIssue{189787} & 1  & Inductor & Wrong Code & Confirmed \\
    & \ptIssue{189799} & 1  & Inductor & Wrong Code & Confirmed \\
    & \ptIssue{189801} & 1  & Inductor & Wrong Code & Confirmed \\
    & \ptIssue{189803} & 1  & Inductor & Wrong Code & Fixed \\
    & \ptIssue{189804} & 1  & Inductor & Wrong Code & Not Planned \\
    & \ptIssue{189808} & 1  & Inductor & Wrong Code & Confirmed \\
    & \ptIssue{190417} & 1  & Inductor & Crash & Confirmed \\
    & \ptIssue{190418} & 1  & Inductor & Crash & Confirmed \\
    & \ptIssue{190421} & 1  & Inductor & Crash & Fixed \\
    & \ptIssue{193581} & 1  & -- & Crash & Reported \\
    & \ptIssue{195963} & 1  & -- & Crash & Confirmed \\
    \midrule
    \multirow{6}{*}{ONNX Runtime}
    & \ortIssue{30513} & 1  & Optimizer & Crash & Confirmed \\
    & \ortIssue{31140} & 1  & CUDA Optimizer & Wrong Code & Reported \\
    & \ortIssue{31161} & 1  & CUDA Optimizer & Wrong Code & Confirmed \\
    & \ortIssue{31162} & 1  & CUDA Optimizer & Wrong Code & Fixed \\
    & \ortIssue{31163} & 1  & CUDA Optimizer & Wrong Code & Reported \\
    & \ortIssue{31980} & 1  & Optimizer & Crash & Confirmed \\
\bottomrule
\end{tabular}
\vspace{2pt}
{\footnotesize \\ $-$Module not yet localized; the crash occurs during
\texttt{torch.compile} but maintainers have not attributed it to a specific pass.}
\end{table*}

Table~\ref{tab:bugs-detail} lists the \numBugs{} distinct bugs that
\toolName{} found in TVM, PyTorch, and ONNX Runtime.
Of these, \numFixed{} have been fixed, \numConfirmed{} have been acknowledged
but not yet fixed, \numReported{} are still awaiting a response, and
\numNotPlanned{} was declined as not planned. Maintainers of all three
independently developed compilers have fixed or acknowledged
\numFixedConfirmed{} of these bugs. Their assessment confirms that those bugs
are genuine faults rather than artifacts of our test generation. Most bugs lie in backend-specific lowering and optimization. GPU scheduling
and CUDA code generation account for 6 of the 8 TVM bugs. Inductor accounts
for all 9 localized PyTorch bugs. The remaining 2 crash during
\texttt{torch.compile}, but maintainers have not yet attributed them to a
specific pass. The ONNX Runtime optimizer accounts for all 6 ONNX Runtime
bugs, 4 of them in its CUDA-specific passes.
The two oracles are complementary. The \numCrashes{} crashes are caught at
compile time. There, an unexpected compiler failure is itself evidence of a
bug, and no reference output is needed. The \numWrongCode{} miscompilations,
by contrast, compile without error and show up only as incorrect output.
It is worth noting that 24 of the 25 bugs were discovered within one month.

\begin{figure*}[!t]
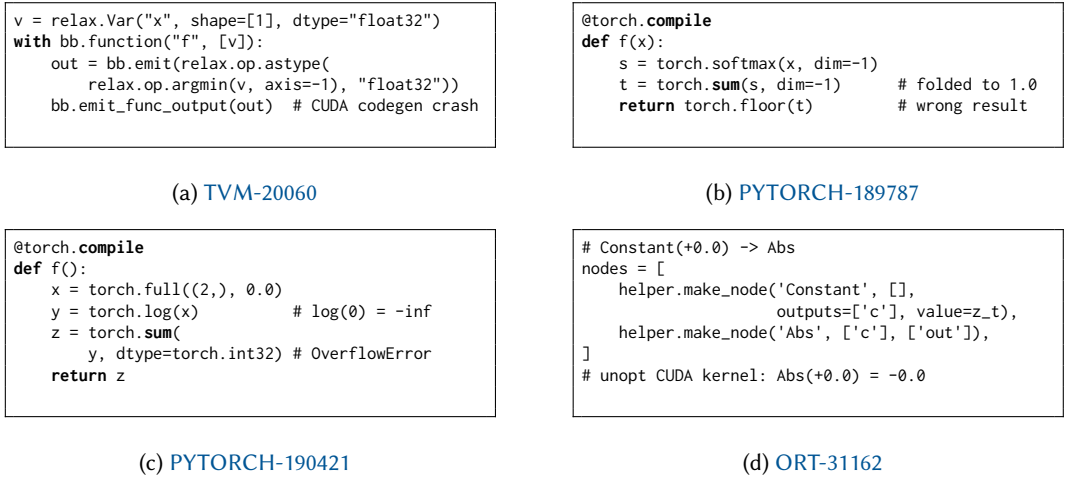

  \centering
  \begin{subfigure}{0.46\linewidth}
    \centering
    \begin{lstlisting}[language=Python,basicstyle=\ttfamily\scriptsize]
v = relax.Var("x", shape=[1], dtype="float32")
with bb.function("f", [v]):
    out = bb.emit(relax.op.astype(
        relax.op.argmin(v, axis=-1), "float32"))
    bb.emit_func_output(out)  # CUDA codegen crash
    \end{lstlisting}
    \caption{ \tvmIssue{20060}}
    \label{fig:bug-tvm}
  \end{subfigure}\hfill
  \begin{subfigure}{0.46\linewidth}
    \centering
    \begin{lstlisting}[language=Python,basicstyle=\ttfamily\scriptsize]
@torch.compile
def f(x):
    s = torch.softmax(x, dim=-1)
    t = torch.sum(s, dim=-1)      # folded to 1.0
    return torch.floor(t)         # wrong result
    \end{lstlisting}
    \caption{\ptIssue{189787}}
    \label{fig:bug-pt-fold}
  \end{subfigure}

  \medskip
  \begin{subfigure}{0.46\linewidth}
    \centering
    \begin{lstlisting}[language=Python,basicstyle=\ttfamily\scriptsize]
@torch.compile
def f():
    x = torch.full((2,), 0.0)
    y = torch.log(x)          # log(0) = -inf
    z = torch.sum(
        y, dtype=torch.int32) # OverflowError
    return z
    \end{lstlisting}
    \caption{ \ptIssue{190421}}
    \label{fig:bug-pt-overflow}
  \end{subfigure}\hfill
  \begin{subfigure}{0.46\linewidth}
    \centering
    \begin{lstlisting}[language=Python,basicstyle=\ttfamily\scriptsize]
# Constant(+0.0) -> Abs
nodes = [
    helper.make_node('Constant', [],
                     outputs=['c'], value=z_t),
    helper.make_node('Abs', ['c'], ['out']),
]
# unopt CUDA kernel: Abs(+0.0) = -0.0
    \end{lstlisting}
    \caption{ \ortIssue{31162}}
    \label{fig:bug-ort}
  \end{subfigure}

  \caption{Four bugs found by \toolName{}: minimal reproducers (a--d) .}
  \label{fig:rq3-bugs}
\end{figure*}
% Preamble: \usepackage{booktabs,tabularx,array}
% and \newcolumntype{Y}[1]{>{\raggedright\arraybackslash\hsize=#1\hsize}X}
\begin{table*}[!t]
\centering
\caption{The semantic graph patterns distilled from the four bugs of
Figure~\ref{fig:rq3-bugs}.}
\label{tab:bug-patterns}
\footnotesize\frenchspacing
\setlength{\tabcolsep}{4pt}
\renewcommand{\arraystretch}{1.15}
\begin{tabularx}{\textwidth}{@{}l Y{1.15} Y{0.80} Y{1.15} Y{0.90}@{}}
\toprule
Bug & (i) Nodes & (ii) Values & (iii) Graph & (iv) Flow \\
\midrule
\tvmIssue{20060}
  & \texttt{n0: argmin, 1$\to$1, axis=-1}\newline
    \texttt{n1: astype, 1$\to$1, dtype=float32}
  & \texttt{x: float32,}\newline\texttt{d[-1] = 1}
  & ---
  & \texttt{n0.out[0] $\to$ n1.in[0]} \\
\addlinespace
\ptIssue{189787}
  & \texttt{n0: softmax, 1$\to$1, dim=-1}\newline
    \texttt{n1: sum, 1$\to$1, dim=-1}\newline
    \texttt{n2: floor, 1$\to$1}
  & \texttt{x: float32,}\newline\texttt{d[-1] $\geq$ 2}
  & ---
  & \texttt{n0.out[0] $\to$ n1.in[0]}\newline
    \texttt{n1.out[0] $\to$ n2.in[0]} \\
\addlinespace
\ptIssue{190421}
  & \texttt{n0: log, 1$\to$1}\newline
    \texttt{n1: sum, 1$\to$1, dtype=int32}
  & \texttt{x: float32}
  & \texttt{x is Constant}
  & \texttt{n0.out[0] $\to$ n1.in[0]} \\
\addlinespace
\ortIssue{31162}
  & \texttt{n0: abs, 1$\to$1}
  & \texttt{x: float32}
  & \texttt{x is Constant; $G$ contains a sign-sensitive operator downstream of n0}
  & --- \\
\bottomrule
\end{tabularx}
\end{table*}

\paragraph{The bugs are diverse.}
The \numBugs{} bugs span three compilers, seven localized modules, and both
crashes and silent miscompilations. Figure~\ref{fig:rq3-bugs} illustrates
this range with four bugs. Together they cover all three compilers, both
symptoms, and faults ranging from graph optimization to code generation.
In Figure~\ref{fig:bug-tvm}, an \texttt{argmin} over a size-1 axis followed by
a dtype cast crashes TVM's CUDA \emph{code generation}.
In Figure~\ref{fig:bug-pt-fold}, an \emph{algebraic rewrite} in PyTorch
Inductor folds $\sum \mathrm{softmax}(x)$ over the same axis to $1.0$. The
identity holds over the reals, but a floating-point sum can fall slightly
below $1.0$. \texttt{floor} then amplifies this rounding difference into
results that differ by a whole integer.
In Figure~\ref{fig:bug-pt-overflow}, a PyTorch graph with no runtime inputs is
evaluated during Inductor's \emph{constant folding}. There, a \texttt{sum}
with \texttt{dtype=int32} converts $\log(0) = -\infty$ to an integer, which
crashes the compiler.
In Figure~\ref{fig:bug-ort}, ONNX Runtime's CUDA \emph{optimizer} processes an
\texttt{abs} applied to a constant, and the optimized model returns $-0.0$
for $|{+0.0}|$, which violates IEEE-754 standard.
The triggers range from ordinary values to degenerate shapes, infinities, and
signed zeros. These four examples are illustrative, not exhaustive. We
quantify diversity over the full set using patterns below.

\paragraph{Patterns behind the bugs.}
Beyond their surface differences, each bug reduces to a compact semantic
graph pattern (Table~\ref{tab:bug-patterns}). A pattern keeps only what is
needed to trigger the bug and abstracts away the rest of the reproducer.
\tvmIssue{20060} is an \texttt{argmin}$\to$\texttt{astype} chain whose input
has a size-1 last axis.
\ptIssue{189787} is a \texttt{softmax}$\to$\texttt{sum}$\to$\texttt{floor}
chain in which \texttt{softmax} and \texttt{sum} reduce over the same axis,
and that axis has at least two elements. A single-element softmax sums to
exactly $1.0$, so the fold would be exact.
\ptIssue{190421} is a \texttt{log}$\to$\texttt{sum} chain in which
\texttt{sum} has \texttt{dtype=int32}. Its input is a constant, so the chain
is folded at compile time.
\ortIssue{31162} is a single \texttt{abs} node on a constant, with a
sign-sensitive operator somewhere downstream that makes the $-0.0$
observable.
The patterns constrain different dimensions. Node and value constraints alone
capture \tvmIssue{20060} and \ptIssue{189787}. In contrast, \ptIssue{190421}
and \ortIssue{31162} also require graph-level constraints: a constant input
for both, and, for \ortIssue{31162}, a downstream sign-sensitive operator.
No single dimension characterizes all four. For example, a pattern language
limited to nodes and values could not express the latter two.

\paragraph{Patterns per bug.}
The \emph{\# of Patterns} column in Table~\ref{tab:bugs-detail} gives the
number of distinct patterns that triggered each bug in our campaign. It
counts only the triggers we observed, so it is a lower bound on how many
ways a bug can be reached.
Twenty-one bugs, including every PyTorch and ONNX Runtime bug, were triggered
by a single pattern. The remaining four are all in TVM's code generation and
GPU scheduling. Each was triggered by 2 to 11 structurally distinct patterns,
and together they account for 19 of the 40 patterns. The most extreme case,
\tvmIssue{20015}, is an LLVM code generation crash reached through 11
patterns.
A plausible explanation is that these components apply the same rules to
broad classes of operators. A single faulty rule can then be reached from
many structurally different graphs. The PyTorch and ONNX Runtime bugs, by
contrast, depend on the semantics of a specific operator combination, such as
the fold in \ptIssue{189787}. However, the component alone does not determine
the count: \tvmIssue{20060}, also a CUDA code generation bug, has a single
pattern.

The pattern count also determines how completely a known bug can be
suppressed. A single-pattern bug is fully suppressed once its pattern is
known. A multi-pattern bug is suppressed only to the extent that its known
patterns cover its triggers, and it may resurface through a new structure.
Conversely, a pattern that is too coarse may also exclude programs that would
trigger a different bug; RQ1 examines such a case.

\textbf{Answer to RQ3.} Despite UIR's simple design, \toolName{} finds real
bugs deep in production DL compilers. To date, \toolName{} has found
\numBugs{} previously unknown bugs across three independently developed
compilers, most of them in backend lowering and optimization.
The bugs are diverse in compiler, component, symptom, and trigger. Each bug
is captured by one or a few compact patterns over UIR's typed operators and
tensors: 21 bugs by a single pattern, and four TVM bugs by 2 to 11 patterns. These
patterns are what guidance uses to suppress known bugs during synthesis, so
that the generation budget goes toward discovering new ones.

\section{Discussion}
\label{sec:discussion}

\textbf{Implications.}
Our results indicate that pattern-guided program synthesis and fuzzing can suppress reports of known defects at a modest throughput cost.
In RQ1, the two patterns targeting \tvmIssue{20036} eliminate all of its crashes ($216 \to 0$).
Because this defect accounts for 85.0\% of the baseline crashes, total crash volume falls by 88.2\% ($254 \to 30$), and by 90.2\% with the full pattern set ($254 \to 25$).
In RQ2, guidance reduces the number of executed tests by 1.0--17.9\%  without a measurable difference in branch coverage relative to the unguided variant.
RQ3 shows that the bugs we found are captured by one or a few compact patterns over UIR, which makes such suppression applicable in practice.
Taken together, these findings suggest that, in a skewed crash distribution, suppressing a small number of dominant defects accounts for most of the defect reports reduction.

\noindent
\textbf{Completeness and over-suppression.}
The effectiveness of suppression depends on both the patterns available and the light-weighted generation algorithm used.
Four TVM bugs in RQ3 were triggered by 2--11 distinct patterns; these are lower bounds because we count only patterns observed during testing.
Similarly, 25 crashes of \tvmIssue{20047} remain in RQ1 even with all four patterns targeting it, showing that the current patterns do not cover all its triggers. On the other hand, an overly broad pattern may also reject programs that trigger different bugs. For example, \tvmIssue{20048} shares a \texttt{conv2d} scheduling subgraph with \tvmIssue{20047}, but is absent from all guided runs. \tvmIssue{20048} is also absent from \textsf{only A}, which has no pattern targeting \tvmIssue{20048}, so the absence of \tvmIssue{20048} may instead result from stochastic generation or limitations of the current generator.

Our initial plan was to integrate pattern guidance into NeuRI.
However, NeuRI employs a backward-generation algorithm that is not
directly supported by our current pattern-matching mechanism.
This highlights a limitation in the generality of the current
implementation rather than an inherent restriction of pattern
guidance.
As future work, we plan to extend the guidance mechanism to support
different graph-generation algorithms, enabling broader evaluation
and reducing dependence on a particular generator.

\section{Threats to Validity}\label{sec:threats}
Fuzzing is stochastic. We repeat each RQ2 configuration five times, but each RQ1 configuration is a single 10{,}000-program run, so RQ1 reports no variance and its small differences (e.g., \tvmIssue{20047}: 36, 30, 26, and 25 reports across configurations) cannot be interpreted. The baseline and guided variants use identical random seeds, but their program sequences diverge after the first pattern rejection, so RQ1 measures end-to-end effects under a fixed generation budget rather than paired outcomes. Counting distinct defects also requires manual root-cause analysis, which may merge distinct defects or split one defect into several. We mitigate this by comparing stack traces and IR node invocation sites, inspecting crash-inducing programs, and having a second author verify the assignments and resolve disagreements through discussion.

The patterns were distilled from reproducers of the defects, so RQ1 shows that known defects can be suppressed, not that patterns generalize to unseen defects or that guidance finds more distinct defects. Suppression may also remove programs that trigger other defects. \tvmIssue{20048} appeared twice in the baseline and in none of the guided runs. Its triggering depends on a particular tensor shape, and its operator structure overlaps with that of \tvmIssue{20047}, so its absence may result from chance, from collateral suppression by overlapping patterns, or both. With two baseline observations, our data cannot distinguish these explanations. We therefore interpret crash reduction as evidence of fewer repeated reports, not as proof that all distinct bug-triggering behaviors are preserved. Validating patterns against reproducers mitigates, but does not eliminate, this risk. Branch coverage is also only a proxy for testing thoroughness.

RQ1 covers TVM (CUDA) and PyTorch Inductor. On ONNX Runtime, both configurations produced zero crashes over 10{,}000 programs (one seed), so it contributes no redundancy measurement, and the ONNX Runtime redundancy (see Section 2) comes from NeuRI's generator. Results may differ across compilers, backends, and generators. Guidance also depends on the available patterns, and our TVM ablation uses only two pattern groups. Longer runs may produce different coverage and crash distributions. Our redundancy metrics consider crashes only and do not capture non-crashing failures such as miscompilations, which account for 10 of the 25 bugs we found.

% \section{Threats to Validity}
% \label{sec:threats}

% Fuzzing outcomes vary across runs. To mitigate this, we repeat each RQ2 configuration five times. In RQ1, baseline and guided variants share identical random seeds, though generated programs diverge following pattern rejection; thus, comparison reflects end-to-end efficiency under a fixed budget.

% Manual root-cause analysis is prone to misgrouping defects. We mitigate this by inspecting crash traces, IR invocation sites, and failure conditions, with two authors independently reviewing and resolving discrepancies. Moreover, pattern generalization risks suppressing rare bugs sharing structural overlap with known patterns (e.g., TVM issue \tvmIssue{20048}). Thus, crash reduction indicates reduced redundant crashes rather than guaranteed preservation of all distinct bugs. RQ1 evaluates TVM with CUDA using two pattern groups; results may vary for other compilers, backends, pattern sets, or execution budgets. Finally, branch coverage is an imperfect proxy for thoroughness, and our redundancy metrics track crashes rather than non-crashing bugs like miscompilations.
\section{Related Work}

\subsection{DL Compiler Fuzzing}

Existing DL compiler fuzzers primarily use program-level diversity to explore
different compiler behaviors. These approaches can be broadly divided into
generation-based and mutation-based techniques. Generation-based fuzzers
construct computational graphs from scratch. Muffin \cite{gu2022muffin}
synthesizes diverse DL models layer by layer as DAGs, while NNSmith
\cite{liu2023nnsmith} uses SMT-based constraint solving to enforce shape and
type consistency. HirGen \cite{ma2023fuzzing} generates structured
high-level IR graphs to target IR optimization passes. More recent approaches
such as OPERA \cite{shen2026opera} and OATest \cite{shen2026oatest} incorporate
data-driven or optimization-aware strategies to exercise complex compiler
transformations. Mutation-based fuzzers instead derive new tests from
existing programs. FreeFuzz \cite{wei2022free} collects API execution traces
from DL frameworks to guide argument fuzzing, while library-to-compiler
migration techniques \cite{shen2025tale} reuse library test suites to
exercise compiler frontends. Despite their different construction mechanisms,
both lines of work primarily rely on program diversity to reach new compiler
behaviors.

\textit{Our goal is different: rather than maximizing program diversity,
\toolName{} targets bug-level diversity.} Its lightweight UIR uses local
operator signatures to enforce shape and type compatibility without global
constraint solving, trading some structural expressiveness for program
generation. Crucially, \toolName{} uses semantic properties available during
synthesis to identify previously discovered bug-triggering patterns and reject
matching candidates before compilation and execution. This \emph{negative
guidance} progressively excludes known failure-inducing regions and redirects
generation toward unexplored bug behavior, allowing \toolName{} to discover
new bugs despite its lightweight generation model.

\subsection{Bug Deduplication}

Diverse test programs can nevertheless trigger the same underlying bug,
making failure deduplication important for reducing the costs of reproduction,
triage, reduction, and reporting. Existing approaches perform deduplication
after failures have been generated and executed. Crash-oriented techniques
use stack signatures, execution traces, or sanitizer logs to group failures
\cite{serebryany2012addresssanitizer,dang2012rebucket,jiang2021igor,chen2013taming},
while recent approaches use richer program and execution features for
non-crashing failures. $D^3$ \cite{yang2023d3} combines static and dynamic
features, BLADE \cite{chen2025blade} learns semantic failure representations,
BugLens \cite{zhou2025bisection} uses commit bisection to distinguish failure
causes, and Donaldson et al. \cite{donaldson2021test} use
semantics-preserving transformations as heuristic failure signatures.
However, these approaches identify redundancy only after test generation and
execution, so redundant tests still incur testing cost. Stack-based signatures
can also be less informative for silent miscompilations and wrong-code bugs
\cite{tambon2024silent,chen2025blade,liu2023nnsmith}.

\textit{\toolName{} instead prevents such redundancy during test generation.}
Each confirmed failure is minimized and distilled into a semantic pattern,
which is matched incrementally during graph synthesis. Candidates matching
known patterns are rejected before compilation and execution, progressively
excluding known failure-inducing regions and directing the testing budget
toward unexplored behavior. Thus, \toolName{} turns previously discovered bugs
into negative guidance, reducing redundant bug rediscovery rather than merely
deduplicating failures after they occur.

\section{Conclusions and Future Work}
\label{sec:conclusion}

Fuzzers for DL compilers waste testing effort, and developers' triage time,
by repeatedly triggering faults that are already known. We presented
\toolName{}, which distills discovered bugs into semantic graph patterns and
regenerates any node that completes a known pattern during synthesis, so
known triggers are avoided before compilation and execution. Against its
unguided variant, patterns for the dominant known defects reduced crash
reports by 90.2\% on TVM (254 to 25) and by 93.8\% on PyTorch Inductor
(594 to 37), with comparable branch coverage and 1.0--17.9\% fewer executed
tests. These results come from one run per configuration and concern the
defects from which the patterns were distilled, so the results show that known
defects can be suppressed at generation time. In a one-month fuzzing,
\toolName{} found \numBugs{} previously unknown bugs across TVM, PyTorch
Inductor, and ONNX Runtime, \numFixed{} of which have been fixed.

In future work, we plan to apply pattern guidance to other graph generators such as NNSmith~\cite{liu2023nnsmith} and NeuRI~\cite{liu2023neuri}, by mapping UIR patterns onto each generator's operators and construction order, to test whether guidance helps generators that already diversify programs; evaluate generalization with repeated runs, patterns applied to fresh seeds, and distinct defects over time against unguided baselines; and automate pattern extraction by generalizing a reduced reproducer one dimension at a time, keeping each relaxation only if matching programs still trigger the same fault. Running this online, as soon as a fault is first reported, would suppress it for the rest of the run.

\section*{Data Availability}
An anonymized replication package is available at~\cite{dataset}.

\bibliographystyle{ACM-Reference-Format}
\bibliography{ref}

@article{chen2025blade,
  author    = {Chen, Junjie and Fan, Xingyu and Yang, Chen and Liu, Shuang and Sun, Jun},
  title     = {De-duplicating Silent Compiler Bugs via Deep Semantic Representation},
  journal   = {Proceedings of the ACM on Software Engineering},
  volume    = {2},
  number    = {FSE},
  pages     =  {2359--2381},
  year      = {2025}
}

@inproceedings{yang2023d3,
  author    = {Yang, Chen and Chen, Junjie and Fan, Xingyu and Jiang, Jiajun and Sun, Jun},
  title     = {Silent Compiler Bug De-duplication via Three-Dimensional Analysis},
  booktitle = {Proceedings of the 32nd ACM SIGSOFT International Symposium on Software Testing and Analysis (ISSTA)},
  pages     = {1122--1134},
  year      = {2023}
}

@inproceedings{zhou2025bisection,
  author    = {Zhou, Xintong and Xu, Zhenyang and Tian, Yongqiang and Sun, Chengnian},
  title     = {On the Feasibility of Deduplicating Compiler Bugs with Bisection},
  booktitle = {Proceedings of the 35th ACM SIGSOFT International Symposium on Software Testing and Analysis (ISSTA)},
  year      = {2026}
}

@inproceedings{liu2023neuri,
  title={Neuri: Diversifying dnn generation via inductive rule inference},
  author={Liu, Jiawei and Peng, Jinjun and Wang, Yuyao and Zhang, Lingming},
  booktitle={Proceedings of the 31st ACM Joint European Software Engineering Conference and Symposium on the Foundations of Software Engineering (ESEC/FSE)},
  pages={657--669},
  year={2023}
}

@inproceedings{liu2023nnsmith,
  author    = {Liu, Jiawei and Lin, Jinkun and Ruffy, Fabian and Tan, Cheng and Li, Jinyang and Panda, Aurojit and Zhang, Lingming},
  title     = {NNSmith: Generating Diverse and Valid Test Cases for Deep Learning Compilers},
  booktitle = {Proceedings of the 28th ACM International Conference on Architectural Support for Programming Languages and Operating Systems (ASPLOS)},
  pages     = {530--543},
  year      = {2023}
}

@inproceedings{gu2022muffin,
  title={Muffin: Testing deep learning libraries via neural architecture fuzzing},
  author={Gu, Jiazhen and Luo, Xuchuan and Zhou, Yangfan and Wang, Xin},
  booktitle={Proceedings of the 44th International Conference on Software Engineering (ICSE)},
  pages={1418--1430},
  year={2022}
}

@inproceedings{shen2025tale,
  author    = {Shen, Qingchao and Ma, Haoyang and Tian, Yongqiang and Chen, Junjie and Cheung, Shing-Chi},
  title     = {A Tale of Two DL Cities: When Library Tests Meet Compiler},
  booktitle = {Proceedings of the 47th IEEE/ACM International Conference on Software Engineering (ICSE)},
  pages     = {2201--2212},
  year      = {2025}
}

@inproceedings{shen2026opera,
  title={Data-driven Test Generation for Fuzzing AI Compiler},
  author={Shen, Qingchao},
  booktitle={Proceedings of the 48th IEEE/ACM International Conference on Software Engineering (ICSE)},
  pages={123--125},
  year={2026}
}

@inproceedings{shen2026oatest,
  title={Optimization-aware test generation for deep learning compilers},
  author={Shen, Qingchao and Wang, Zan and Ma, Haoyang and Tian, Yongqiang and Huang, Lili and Xiao, Zibo and Chen, Junjie and Cheung, Shing-Chi},
  booktitle={Proceedings of the 48th IEEE/ACM International Conference on Software Engineering (ICSE)},
  pages={3787--3799},
  year={2026}
}

@inproceedings{serebryany2012addresssanitizer,
  title={$\{$AddressSanitizer$\}$: A fast address sanity checker},
  author={Serebryany, Konstantin and Bruening, Derek and Potapenko, Alexander and Vyukov, Dmitriy},
  booktitle={Proceedings of the 30th USENIX Annual Technical Conference (USENIX ATC)},
  pages={309--318},
  year={2012}
}

@inproceedings{dang2012rebucket,
  author    = {Dang, Yingnong and Wu, Rongxin and Zhang, Hongyu and Zhang, Dongmei and Nobel, Peter},
  title     = {ReBucket: A Method for Clustering Duplicate Crash Reports Based on Call Stack Similarity},
  booktitle = {Proceedings of the 34th International Conference on Software Engineering (ICSE)},
  pages     = {1084--1093},
  year      = {2012}
}

@inproceedings{jiang2021igor,
  title={Igor: Crash deduplication through root-cause clustering},
  author={Jiang, Zhiyuan and Jiang, Xiyue and Hazimeh, Ahmad and Tang, Chaojing and Zhang, Chao and Payer, Mathias},
  booktitle={Proceedings of the 28th ACM SIGSAC Conference on Computer and Communications Security (CCS)},
  pages={3318--3336},
  year={2021}
}

@article{tambon2024silent,
  title={Silent bugs in deep learning frameworks: an empirical study of keras and tensorflow},
  author={Tambon, Florian and Nikanjam, Amin and An, Le and Khomh, Foutse and Antoniol, Giuliano},
  journal={Empirical Software Engineering},
  volume={29},
  number={1},
  pages={1--34},
  year={2024},
  publisher={Springer}
}

@inproceedings{wei2022free,
  title={Free lunch for testing: Fuzzing deep-learning libraries from open source},
  author={Wei, Anjiang and Deng, Yinlin and Yang, Chenyuan and Zhang, Lingming},
  booktitle={Proceedings of the 44th International Conference on Software Engineering (ICSE)},
  pages={995--1007},
  year={2022}
}

@inproceedings{chen2013taming,
  title={Taming compiler fuzzers},
  author={Chen, Yang and Groce, Alex and Zhang, Chaoqiang and Wong, Weng-Keen and Fern, Xiaoli and Eide, Eric and Regehr, John},
  booktitle={Proceedings of the 34th ACM SIGPLAN Conference on Programming Language Design and Implementation (PLDI)},
  pages={197--208},
  year={2013}
}

@inproceedings{donaldson2021test,
  title={Test-case reduction and deduplication almost for free with transformation-based compiler testing},
  author={Donaldson, Alastair F and Thomson, Paul and Teliman, Vasyl and Milizia, Stefano and Maselco, Andr{\'e} Perez and Karpi{\'n}ski, Antoni},
  booktitle={Proceedings of the 42nd ACM SIGPLAN International Conference on Programming Language Design and Implementation (PLDI)},
  pages={1017--1032},
  year={2021}
}

@inproceedings{tvm,
  title={TVM: An automated End-to-End optimizing compiler for deep learning},
  author={Chen, Tianqi and Moreau, Thierry and Jiang, Ziheng and Zheng, Lianmin and Yan, Eddie and Shen, Haichen and Cowan, Meghan and Wang, Leyuan and Hu, Yuwei and Ceze, Luis and others},
  booktitle={Proceedings of the 13th USENIX Symposium on Operating Systems Design and Implementation (OSDI)},
  pages={578--594},
  year={2018}
}

@inproceedings{csmith,
  title={Finding and understanding bugs in C compilers},
  author={Yang, Xuejun and Chen, Yang and Eide, Eric and Regehr, John},
  booktitle={Proceedings of the 32nd ACM SIGPLAN Conference on Programming Language Design and Implementation (PLDI)},
  pages={283--294},
  year={2011}
}

@article{afl,
  title={Technical “whitepaper” for afl-fuzz},
  author={Zalewski, Michal},
  journal={\url{http://lcamtuf.coredump.cx/afl/technical details.txt}},
  year={2014}
}

@misc{torchinductor-patternmatcher,
  title        = {Pattern Matcher and Building a Convolution/Batch Norm Fuser with torch.compile},
  author       = {{PyTorch}},
  howpublished = {\url{https://docs.pytorch.org/tutorials/intermediate/torch_compile_conv_bn_fuser.html}},
  note         = {See also \texttt{torch.\_inductor.pattern\_matcher} in the PyTorch source: \url{https://github.com/pytorch/pytorch/blob/main/torch/_inductor/pattern_matcher.py}},
  organization = {PyTorch Foundation}
}

@misc{tvm-fuseops,
  author       = {{Apache TVM}},
  title        = {Operator Fusion in {Relax}: the \texttt{FuseOps} and \texttt{FuseTIR} Passes},
  howpublished = {\url{https://github.com/apache/tvm/blob/main/src/relax/transform/fuse_ops.cc}},
  note         = {Fusion categories given by the \texttt{OpPatternKind} enumeration
                  (\texttt{kElemWise}, \texttt{kBroadcast}, \texttt{kInjective},
                  \texttt{kCommReduce}, \texttt{kOutEWiseFusable}, \texttt{kTuple},
                  \texttt{kOpaque}); accessed Sept.\ 2026},
  year         = {2026}
}

@inproceedings{hicond,
  title={History-guided configuration diversification for compiler test-program generation},
  author={Chen, Junjie and Wang, Guancheng and Hao, Dan and Xiong, Yingfei and Zhang, Hongyu and Zhang, Lu},
  booktitle={Proceedings of the 34th IEEE/ACM International Conference on Automated Software Engineering (ASE)},
  pages={305--316},
  year={2019}
}

@inproceedings{swarm,
  title={Swarm testing},
  author={Groce, Alex and Zhang, Chaoqiang and Eide, Eric and Chen, Yang and Regehr, John},
  booktitle={Proceedings of the 21st International Symposium on Software Testing and Analysis (ISSTA)},
  pages={78--88},
  year={2012}
}

@inproceedings{klees2018evaluating,
  title={Evaluating fuzz testing},
  author={Klees, George and Ruef, Andrew and Cooper, Benji and Wei, Shiyi and Hicks, Michael},
  booktitle={Proceedings of the 25th ACM SIGSAC Conference on Computer and Communications Security (CCS)},
  pages={2123--2138},
  year={2018}
}

@inproceedings{pytorch,
  title={Pytorch 2: Faster machine learning through dynamic python bytecode transformation and graph compilation},
  author={Ansel, Jason and Yang, Edward and He, Horace and Gimelshein, Natalia and Jain, Animesh and Voznesensky, Michael and Bao, Bin and Bell, Peter and Berard, David and Burovski, Evgeni and others},
  booktitle={Proceedings of the 29th ACM International Conference on Architectural Support for Programming Languages and Operating Systems (ASPLOS)},
  pages={929--947},
  year={2024}
}

@misc{onnxruntime,
  author       = {{ONNX Runtime developers}},
  title        = {{ONNX} {Runtime}},
  howpublished = {\url{https://onnxruntime.ai}},
  year         = {2021},
  note         = {Version 1.27.0; accessed Sept.\ 2026}
}

@article{zhu1997coverage,
  title={Software unit test coverage and adequacy},
  author={Zhu, Hong and Hall, Patrick AV and May, John HR},
  journal={ACM Computing Surveys},
  volume={29},
  number={4},
  pages={366--427},
  year={1997},
  publisher={ACM}
}

@inproceedings{ma2023fuzzing,
  title={Fuzzing deep learning compilers with hirgen},
  author={Ma, Haoyang and Shen, Qingchao and Tian, Yongqiang and Chen, Junjie and Cheung, Shing-Chi},
  booktitle={Proceedings of the 32nd ACM SIGSOFT International Symposium on Software Testing and Analysis (ISSTA)},
  pages={248--260},
  year={2023}
}

@inproceedings{wang2023gencog,
  title={Gencog: A dsl-based approach to generating computation graphs for tvm testing},
  author={Wang, Zihan and Nie, Pengbo and Miao, Xinyuan and Chen, Yuting and Wan, Chengcheng and Bu, Lei and Zhao, Jianjun},
  booktitle={Proceedings of the 32nd ACM SIGSOFT International Symposium on Software Testing and Analysis (ISSTA)},
  pages={904--916},
  year={2023}
}

@article{liu2022coverage,
  title={Coverage-guided tensor compiler fuzzing with joint ir-pass mutation},
  author={Liu, Jiawei and Wei, Yuxiang and Yang, Sen and Deng, Yinlin and Zhang, Lingming},
  journal={Proceedings of the ACM on Programming Languages},
  volume={6},
  number={OOPSLA1},
  pages={1--26},
  year={2022},
  publisher={ACM}
}

@article{yang2025may,
  title={May the feedback be with you! unlocking the power of feedback-driven deep learning framework fuzzing via LLMs},
  author={Yang, Shaoyu and Fang, Chunrong and Lin, Haifeng and Chen, Xiang and Liu, Jia and Chen, Zhenyu},
  journal={arXiv preprint arXiv:2506.17642},
  year={2025}
}

@article{simpson1949measurement,
  title={Measurement of diversity},
  author={Simpson, Edward H},
  journal={Nature},
  volume={163},
  number={4148},
  pages={688--688},
  year={1949},
  publisher={Nature Publishing Group}
}

@article{hill1973diversity,
  author  = {Hill, Mark O.},
  title   = {Diversity and Evenness: A Unifying Notation and Its Consequences},
  journal = {Ecology},
  volume  = {54},
  number  = {2},
  pages   = {427--432},
  year    = {1973}
}

@article{jost2006entropy,
  title={Entropy and diversity},
  author={Jost, Lou},
  journal={Oikos},
  volume={113},
  number={2},
  pages={363--375},
  year={2006},
  publisher={Wiley}
}

@article{feld1977classsize,
  author  = {Feld, Scott L. and Grofman, Bernard},
  title   = {Variation in Class Size, the Class Size Paradox, and Some
             Consequences for Students},
  journal = {Research in Higher Education},
  volume  = {6},
  pages   = {215--222},
  year    = {1977}
}

@article{bohme2018stads,
  title={STADS: Software testing as species discovery},
  author={B{\"o}hme, Marcel},
  journal={ACM Transactions on Software Engineering and Methodology},
  volume={27},
  number={2},
  pages={1--52},
  year={2018},
  publisher={ACM}
}

@misc{dataset,
  author = {Qiong Feng and Xiaotian Ma and Di Cui and Wei Song and Peng Liang},
  title  = {Replication Package for ``Pattern-Guided Graph Synthesis for Suppressing Known Defects in DL Compiler Fuzzing''},
  year   = {2026},
  url    = {}
}

\end{document}